\documentclass[11pt,a4paper]{article}
\usepackage{jheppub}
\usepackage[utf8]{inputenc}
\usepackage{graphicx,fancybox,float,comment}

\usepackage{units}
\usepackage{subcaption}
\usepackage{multirow,array,arydshln}
\usepackage[dvipsnames]{xcolor}
\usepackage{braket,tensor}
\usepackage{amsmath,amssymb,amsfonts,ulem}
\usepackage{tikz-cd}

\newcommand{\be}{\begin{equation}}
\newcommand{\ee}{\end{equation}}

\global\interfootnotelinepenalty=1000

\graphicspath{{graphics}}

\title{Horizon symmetries of exact hydrodynamic actions}

\author[a]{Mike Blake,}
\author[b,c,d]{Arpit Das}
\author[e]{and Richard A. Davison}

\affiliation[a]{School of Mathematics, University of Bristol, Woodland Road, Bristol BS8 1UG, U.K.} 
\affiliation[b]{School of Mathematics and Maxwell Institute for Mathematical Sciences, University of Edinburgh,\\ Edinburgh EH9 3FD, U.K.}
\affiliation[c]{Higgs Centre for Theoretical Physics,
University of Edinburgh,\\ Edinburgh EH8 9YL, U.K.}
\affiliation[d]{Department of Astrophysics and High Energy Physics, S.N. Bose National Centre for Basic Sciences, Salt Lake, Kolkata 700106, India}
\affiliation[e]{Department of Mathematics and Maxwell Institute for Mathematical Sciences,\\ Heriot-Watt University, Edinburgh EH14 4AS, U.K.}

\emailAdd{mike.blake@bristol.ac.uk}
\emailAdd{adas95@bose.res.in}
\emailAdd{r.davison@hw.ac.uk}

\abstract{We demonstrate how horizon symmetries of AdS black holes are realised in the hydrodynamic actions for the dynamics of the dual stress tensor. A recent approach for obtaining such actions involves a choice of horizon boundary conditions, and we show that diffeomorphisms that preserve these give rise to symmetries of the corresponding hydrodynamic action. We obtain explicit hydrodynamic actions for small amplitude perturbations of black holes in JT and AdS$_3$ gravity using this approach and demonstrate explicitly the existence of these symmetries. For a natural choice of boundary conditions, the resulting actions manifestly realise gauge symmetries previously proposed for effective theories of maximal chaos.}

\begin{document}

\maketitle

\section{Introduction}

\paragraph{}The macroscopic thermal dynamics of the stress tensor in a quantum field theory is governed by the theory of hydrodynamics. In recent years it has been established that an action principle for this effective theory can be constructed, where the degrees of freedom are diffeomorphism fields acting as maps between a fluid spacetime and two physical spacetimes \cite{Crossley:2015evo,Liu:2018kfw} (see also \cite{Grozdanov:2013dba,Haehl:2015foa,Haehl:2017zac,Jensen:2017kzi} for other approaches). Such actions have led to significant progress. For instance in constraining the form of constitutive relations from first principles, calculating hydrodynamic loop effects, and proposing connections with quantum chaos (see \cite{Delacretaz:2026owo} for a recent overview). 

\paragraph{} A key motivation in these developments has been the study of holographic quantum field theories, which have a simpler description as theories of gravity. In these theories, the hydrodynamic degrees of freedom are believed to arise naturally as relative diffeomorphisms between the horizon and the asymptotic boundaries of maximally extended anti-de Sitter space time \cite{Nickel:2010pr,Crossley:2015tka,deBoer:2015ija,Glorioso:2018mmw,Liu:2018crr} (see also \cite{deBoer:2018qqm} for the hydrodynamics of a conserved U(1) current). In the recent work \cite{Blake:2026jue} an explicit prescription was proposed that incorporates dissipative effects in the hydrodynamic action, working to quadratic order in perturbations around the thermal state (see \cite{Jana:2020vyx,Ghosh:2020lel,He:2021jna,He:2022jnc,He:2022deg,Loganayagam:2022zmq,Loganayagam:2022teq} for a complementary approach, and \cite{Iqbal:2020lrt,Loganayagam:2020eue,Loganayagam:2020iol,Bu:2021clf,Bu:2022esd,Ghosh:2022fyo,Baggioli:2023tlc,Rangamani:2023mok,Martin:2024mdm,Baggioli:2024zfq,Bu:2024fhz,Ammon:2025vod,Bu:2026mxr} for related holographic studies of other operators).\footnote{See \cite{Blake:2026jue} for a detailed discussion of the alternative proposal of \cite{Bu:2025zad}.}

\paragraph{}There are special cases of holographic systems in which exact hydrodynamic actions can be derived directly from gravity. One is the Schwarzian action of 2-dimensional JT gravity \cite{Maldacena:2016upp,Engelsoy:2016xyb,Jensen:2016pah} and another is pure gravity in AdS$_3$ \cite{Cotler:2018zff,Nguyen:2021pdz} (see also \cite{Haehl:2018izb,Haehl:2019eae,Nguyen:2021jja, Nguyen:2022xsw,Nguyen:2025jhc} for further perspectives on this). In the gravitational description, these theories are special as there are only boundary dynamics. This reflects the fact that the field theory stress tensor is highly constrained by symmetries: for example, its two-point functions are completely determined by the Ward identities for diffeomorphisms and Weyl transformations. Because of this, they are especially simple examples of hydrodynamic theories which are non-dissipative and for which the `effective' theory is actually valid at all scales.

\paragraph{}Another advantage of these special systems is that their quantum chaotic properties are readily accessible (see \cite{Maldacena:2016hyu,Roberts:2014ifa} in addition to the references above, as well as \cite{Gu:2016oyy,Choi:2020tdj,Bucca:2026xmh} for spatially extended generalizations of the Schwarzian action). In particular, generic out-of-time-ordered correlators in these systems have a characteristic form that is closely related to SL(2,$\mathbb{R}$) symmetries in the hydrodynamic action. In \cite{Blake:2017ris,Blake:2021wqj} it was conjectured that this was a simple manifestation of a much more general phenomenon: that the universal form of generic out-of-time-ordered correlators in holographic theories is inherited from characteristic symmetries in their hydrodynamic actions.\footnote{See \cite{Grozdanov:2017ajz,Blake:2018leo} for indirect tests of this proposal.} It was further conjectured in \cite{Knysh:2024asf} that in the gravitational description these symmetries are particular geometric symmetries of black hole horizons (see \cite{Eling:2016qvx} for a previous, related proposal).

\paragraph{}In this paper we revisit the construction of hydrodynamic actions for JT and AdS$_3$ gravity, following the approach introduced in \cite{Blake:2026jue}.\footnote{More precisely we will use a simplified version of the method of \cite{Blake:2026jue} that uses only a single black hole exterior. This simplification is possible for these theories as their momentum space retarded and advanced stress tensor two-point functions are real (up to an $i\epsilon$ prescription) and so can be obtained from a hydrodynamic theory with a single physical spacetime.} The motivation for this is two-fold. The first is as a pedagogical demonstration, and consistency check, of the approach of \cite{Blake:2026jue}. In particular, the non-effective nature of the hydrodynamic theory in these examples means that the approach of \cite{Blake:2026jue} can be implemented exactly (i.e.~with no derivative expansion) to produce a hydrodynamic action whose degrees of freedom are relative diffeomorphisms between the asymptotic boundary and horizon of the spacetime. Secondly we will demonstrate that these hydrodynamic actions inherit specific gauge symmetries that can be understood directly as black hole horizon symmetries, explicitly realising the idea proposed in \cite{Knysh:2024asf}.

\paragraph{}The hydrodynamic action is obtained by evaluating the gravitational action (exterior to the black hole) for partially on-shell configurations of the fields. This requires a choice of horizon boundary conditions, upon which the action and its symmetries ultimately depend. For mixed boundary conditions analogous to those studied in higher dimensions \cite{Blake:2026jue}, we find that the horizon symmetries of the hydrodynamic actions of JT and AdS$_3$ gravity are precisely those proposed in \cite{Knysh:2024asf}. For JT gravity with these boundary conditions the action is the quadratic Schwarzian for relative diffeomorphisms, and so we obtain a new geometric understanding of its known symmetries in terms of horizon physics. For AdS$_3$ gravity with these boundary conditions we find a quadratic action for relative diffeomorphisms that is different from those of \cite{Haehl:2018izb,Cotler:2018zff,Nguyen:2021jja,Nguyen:2022xsw},\footnote{As we will explain, our action is related to one of those obtained in \cite{Nguyen:2021pdz}.} but that still produces the thermal two-point functions of a 2D CFT stress tensor.

\paragraph{}The horizon symmetries of the hydrodynamic action manifestly depend upon the choice of horizon boundary conditions. We demonstrate this further by also deriving the hydrodynamic actions obtained upon imposing Dirichlet boundary conditions on the (stretched) horizon. These are equally valid quadratic actions (i.e.~they produce the same two-point functions at generic frequencies)  with horizon symmetries different than those of \cite{Knysh:2024asf}.

\section{Horizon symmetries} \label{sec:hor_sy}

\paragraph{}We begin by explaining what we mean by horizon symmetries, and how they arise in hydrodynamic actions. Horizon symmetries are diffeomorphisms that preserve a set of near-horizon boundary conditions for a spacetime metric. In this sense they are somewhat similar to asymptotic symmetries. We will show that these diffeomorphisms act non-trivially on the degrees of freedom of the hydrodynamic action for the dual field theory. This was anticipated in~\cite{Knysh:2024asf} and one of the key goals of this paper is to explicitly demonstrate this using the examples of JT and AdS$_3$ gravity.

\paragraph{} Our starting point is a $d$-dimensional, asymptotically AdS equilibrium black brane geometry with metric
\begin{equation}
ds^2 = -D(r) dt^2 + 2 dt dr + r^2 dx^{i} dx^{i}, \nonumber 
\end{equation}
where $i=1,\ldots,d-2$ are spatial dimensions and $D(r)$ has a simple zero at the horizon $r=r_0$. Note that $t$ is an ingoing Eddington-Finkelstein-like coordinate. By evaluating the gravitational action for partially on-shell configurations (up to quadratic order in perturbations around equilibrium), one obtains a hydrodynamic action whose dynamical degrees of freedom $\xi_\mu(t,x^i)$ are relative diffeomorphisms between the asymptotic boundary and the horizon.\footnote{We use indices $\mu,\nu,\ldots$ to denote boundary coordinates $(t,x^i)$.}

\paragraph{}In this paper we will consider two theories for which the hydrodynamics is non-dissipative, and for which $D(r)=r^2-r_0^2$. For JT gravity ($d=2$) there is a single hydrodynamical field
\begin{equation}
\xi_t(t) = [-\zeta^{t}(t, r)]^{r=\infty}_{r=r_0},
\label{hydrodefs2}
\end{equation}
where in general $\zeta^M(t,r,x^i)$ denotes a diffeomorphism that appears naturally in our parameterisation of the perturbation of the spacetime metric.\footnote{We use indices $M,N,\ldots$ to denote bulk coordinates $(r,t,x^i)$.} For AdS$_3$ gravity ($d=3$) there are two hydrodynamical fields
\begin{equation}
\xi_t(t,x) = [-\zeta^{t}(t,r,x)]^{\infty}_{r_0}, \hspace{2.0cm}  \xi_x(t, x)= [\zeta^{x}(t,r,x) - \partial_{x} \zeta^{t}(t,r,x)/r]^{\infty}_{r_0}. 
\label{hydrodefs3}
\end{equation}
For a generic theory the hydrodynamics will be dissipative and so will involve a non-trivial coupling between two copies of the hydrodynamical fields (one for each branch of the Schwinger-Keldysh contour). Each copy is a set of relative diffeomorphisms between the horizon and the asymptotic boundary on one branch of the Crossley-Glorioso-Liu contour (see \cite{Blake:2026jue} for the $d=4$ case which is of this kind).

\paragraph{}To obtain an action for the hydrodynamical fields, we choose a set of horizon boundary conditions for the partially on-shell configurations. As we will see later, the form of the action, and its symmetries, depend on this choice. We can abstractly denote the horizon boundary conditions by the equation
\begin{equation}
    \mathcal{C}_A[\Psi]=0,
\label{eq:general-horizon-bcs}
\end{equation}
at $r=r_0$, where $\Psi$ denotes the bulk field perturbations and $\mathcal{C}_A$ are bulk differential operators (i.e.~they can include radial derivatives) indexed by the label $A$. Consider a bulk diffeomorphism $\chi^M(t,r,x^i)$ that is trivial near the asymptotic boundary but acts non-trivially near the horizon as
\begin{equation}
\Psi \to \Psi + \delta \Psi, \hspace{2.0cm} \delta \Psi = {\mathcal L}_{\chi} \Psi,
\end{equation}
where $\mathcal{L}$ is the Lie derivative. We call this diffeomorphism a horizon symmetry if it preserves the horizon boundary conditions \eqref{eq:general-horizon-bcs} i.e.~if
\begin{equation}
    \mathcal C_A[\delta \Psi]
 =0 ,
    \label{eq:general-horizon-syms}
\end{equation}
at the horizon.

\paragraph{}The bulk diffeomorphism $\chi^M$ induces shifts of the hydrodynamical degrees of freedom $\xi_\mu$, since it shifts the horizon values of $\zeta^M$ (recall equations~\eqref{hydrodefs2} and~\eqref{hydrodefs3}). Specifically, if we follow the notation of \cite{Knysh:2024asf} and parameterise the diffeomorphism as
\begin{equation}
\chi(t, r \to r_0, x^{i})  = f(t, x^{i}) \left(\frac{\partial}{\partial t}\right) + Z(t, x^{i})\left(\frac{\partial}{\partial r}\right) + \tilde{X}^{i}(t,x^{i}) \left(\frac{\partial}{\partial x^{i}}\right) + \dots, 
\label{diff2}
\end{equation}
then for JT gravity this shifts the hydrodynamic field \eqref{hydrodefs2} by
\begin{equation}
    \delta_{H} \xi_t(t)  = f(t),
\end{equation}
and for AdS$_3$ gravity it shifts the hydrodynamic fields \eqref{hydrodefs3} by
\begin{equation}
    \delta_{H} \xi_t(t, x)  = f(t,x ), \hspace{2.0cm}  \delta_{H} \xi_x(t,x)  = -\tilde{X}(t,x) + \partial_{x} f(t,x)/r_0.
\end{equation}
The diffeomorphism $\chi^M$ is only a horizon symmetry for specific choices of the functions $f$, $Z$ and $\tilde{X}$. These are the diffeomorphisms which are unbroken by the horizon boundary conditions and as such we expect them to be realised as gauge symmetries of the hydrodynamic action. We will explicitly demonstrate this by constructing hydrodynamic actions for JT and AdS$_3$ gravity. More precisely we will show that the hydrodynamic action, solutions to its equations of motion, and its constitutive relations, are all invariant under the shifts induced by horizon symmetries as defined here.

\subsection*{Horizon boundary conditions}

\paragraph{}The preceding definition of horizon symmetries requires a choice of horizon boundary conditions. One natural choice is Dirichlet boundary conditions on the field perturbations $\Psi$ on the stretched horizon, which requires the addition of a Gibbons-Hawking term to render the variational principle well-defined. Here we will instead consider the following mixed set of boundary conditions for the perturbations of the spacetime metric $\delta g_{MN}$
\begin{equation}
    \left.\delta g_{t\mu}\right|_{r_0}=0,
    \qquad
    \left.
       \partial_r\delta g_{tt}
       -2\partial_t\delta g_{tr}
       +2\kappa_0\delta g_{tr}
    \right|_{r_0}
    =0 ,
    \label{ads23bcsi}
\end{equation}
where $\kappa_0 = D'(r_0)/2=2\pi T$ is the equilibrium surface gravity (and $T$ is the equilibrium temperature).

\paragraph{}The boundary conditions \eqref{ads23bcsi} are motivated dynamically -- for the case of JT and AdS$_3$ gravity they ensure that the variational principle is satisfied in the absence of a Gibbons-Hawking term at the horizon (see Appendix \ref{gen_bcs_JT-new}). Consistency with the variational principle ensures that the equations of motions of the resulting hydrodynamic action are equivalent to the Einstein equations that were not yet imposed on the partially on-shell field configurations.

\paragraph{} It is straightforward to determine the horizon symmetries that preserve these boundary conditions. Requiring that~\eqref{ads23bcsi} are preserved fixes the form of the near-horizon diffeomorphism~\eqref{diff2} to be 
\begin{eqnarray}
f(t,x^{i}) &=& a_1(x^{i}) + a_2(x^{i}) e^{-\kappa_0 t} - \frac{D''(r_0)}{4 \kappa_0^2} a_3(x^{i}) e^{\kappa_0 t}, \nonumber \\
Z(t, x^{i}) &=& a_3(x^{i}) e^{\kappa_0 t}, \nonumber \\
\tilde{X}^{i}(t,x^{i}) &=& a_{4}^{i}(x^{i}) - \frac{1}{\kappa_0 r_0^2} \partial_{x^{i}} a_3(x^{i}) e^{\kappa_0 t}. 
\label{hongsymmetry} 
\end{eqnarray} 
These expressions are the same as those obtained in~\cite{Knysh:2024asf}, which defined horizon symmetries as diffeomorphisms that preserve certain geometric quantities at the horizon. A subset of these symmetries have previously been referred to as supertranslations and superrotations \cite{Donnay:2015abr,Donnay:2016ejv,Chandrasekaran:2018aop,Marjieh:2021gln}. Our discussion emphasises a new perspective on them: for perturbations of a black brane they are diffeomorphisms that preserve boundary conditions arising naturally from the variational principle of JT and AdS$_3$ gravity (in the absence of a horizon Gibbons-Hawking term). Furthermore, whilst~\cite{Knysh:2024asf} argued that these should be realised as gauge symmetries of the hydrodynamic action, it did not construct hydrodynamic actions that realise them. The main purpose of this paper is to explicitly do this.

\paragraph{}For the case of JT gravity, the horizon symmetries \eqref{hongsymmetry} induce the following shift of the hydrodynamic field
\begin{equation}
\delta_H\xi_t(t) = a_1 + a_2 e^{- 2\pi T t} + \bar{a}_3 e^{2\pi T t},
\end{equation}
where $\bar{a}_3 = -a_3/(2 r_0^2)$, which are the familiar zero modes of the Schwarzian action. Indeed we will see that the hydrodynamic action computed using the boundary conditions~\eqref{ads23bcsi} is precisely the quadratic Schwarzian action for the relative diffeomorphism $\xi_t$, coupled to external sources. Whilst this may seem unsurprising to the reader familiar with JT gravity, we note that from our perspective it is a non-trivial consequence of the boundary conditions~\eqref{ads23bcsi}. For AdS$_3$ gravity the horizon symmetries induce the following shifts
\begin{equation}
\begin{aligned}
\delta_H\xi_t (t,x)&\,= a_1(x) +  a_2(x) e^{- 2\pi T t} +  \bar{a}_3(x) e^{2\pi T t},\\
\delta_H\xi_x(t,x)&\,= a_{x}(x) +  \frac{\partial_x a_2(x)}{r_0} e^{-2\pi T t} - \frac{\partial_x \bar{a}_3(x)}{r_0} e^{2\pi T t},
\end{aligned}
\end{equation}
where $\bar a_3(x)=-a_3(x)/(2r_0^2)$ and $a_x(x)=-a_4^x(x)+\partial_xa_1(x)/r_0$. We will explicitly obtain an action realising such symmetries as zero modes. This action has not, to our knowledge, previously appeared in the literature.

\paragraph{}In Appendix \ref{app:Dir} we consider instead imposing Dirichlet boundary conditions on the stretched horizon, after adding a Gibbons-Hawking term there. One can still identify horizon symmetries of these actions as above: they no longer include pointwise exponentially growing or decaying modes, emphasising that even qualitative properties of the horizon symmetries are dependent on the choice of boundary conditions. Specifically, in the JT case, the resulting action is no longer the Schwarzian, although it still produces the correct retarded stress tensor two-point functions at generic frequencies when the hydrodynamical fields are put on-shell. 

\paragraph{}For a generic (dissipative) theory there are two copies of each field, one on each branch of the Crossley-Glorioso-Liu contour that circles the horizon. In addition to the horizon boundary conditions imposed on the fields on a single branch (analogous to those in \eqref{eq:general-horizon-bcs} above), there are additional horizon boundary conditions that couple the fields on the two branches. These additional conditions (partially) break the horizon symmetries of the hydrodynamic action from two independent copies of \eqref{eq:general-horizon-syms} down to a smaller set. For Schwarzschild-AdS$_4$ \cite{Blake:2026jue} the natural generalisation of the horizon boundary conditions \eqref{ads23bcsi} are not invariant under the horizon symmetries proposed in \cite{Knysh:2024asf}.

\paragraph{}We now turn to explicitly computing the hydrodynamic actions and studying their symmetries. In Section \ref{sec:AdS2section} we do this for JT gravity and in Section \ref{sec:AdS3section} we do it for AdS$_3$ gravity before concluding with a discussion in Section \ref{sec:disc}. Further technical details can be found in the Appendices, as can the actions with Dirichlet boundary conditions (Appendix \ref{app:Dir}).

\section{Hydrodynamic action for JT gravity}
\label{sec:AdS2section}

\paragraph{}In this Section we use the approach of \cite{Blake:2026jue} to derive a hydrodynamic action for the field theory dual to JT gravity. In spirit this is similar to the approach of \cite{Maldacena:2016upp}, in which the boundary Schwarzian action for a boundary time reparameterisation was obtained by evaluating the gravitational action partially on-shell (specifically, after imposing the dilaton's equation of motion), although the details differ. We will still obtain the Schwarzian action, but for a relative time diffeomorphism between the boundary and horizon, working directly in Lorentzian signature.\footnote{We will only work to quadratic order in perturbations around the thermal state, unlike \cite{Maldacena:2016upp} in which the non-linear Schwarzian action is obtained.} While in \cite{Maldacena:2016upp} the bulk metric is gauge fixed and the boundary embedding is dynamical, we will fix the boundary embedding but let the bulk metric vary (which is the approach used in \cite{Blake:2026jue} for black holes with non-trivial bulk dynamics).\footnote{We emphasise that we are not attempting to derive a holographic action for 2-dimensional quantum gravity, but rather than to obtain a large$-N$ action for the boundary stress tensor.}

\subsubsection*{Equilibrium state} 

\paragraph{} Our starting point is the action
\begin{equation}
S_{\mathrm{EH}} = \int d^2x \sqrt{-g}{ \Phi(R + 2)},  
\label{ads2action}
\end{equation}
supplemented with the boundary term\footnote{We are neglecting the anomaly counterterm identified in \cite{Cvetic:2016eiv}, which is a total derivative.}
\begin{equation}\label{eq:bd_JT}
 S_{\mathrm{GH}} =2 \int_{\mathrm{boundary}}  dt \sqrt{-\gamma}\Phi (K-1). 
\end{equation}
We will refer to these as the Einstein-Hilbert and Gibbons-Hawking pieces of the action (even though they both contain a dilaton, and the latter also contains a counterterm). We will denote the full action by $S_{\mathrm{JT}} = S_{\mathrm{EH}} + S_{\mathrm{GH}}$. The dilaton is dual to a scalar operator in the field theory, which we denote $\mathcal{O}_\phi$.

\paragraph{} The equations of motion obtained from varying the metric are given by 
\begin{equation}
E_{M N} = -\nabla_{M} \nabla_{N} \Phi + g_{M N} \nabla^2 \Phi - g_{M N} \Phi = 0,
\label{metriceom}
\end{equation}
while those obtained by varying with respect to the dilaton are
 \begin{equation}
 E_{\Phi} = R + 2 = 0.
 \label{scalareom}
 \end{equation} 
\paragraph{} We will work in coordinates where the asymptotic boundary lies at $r \to \infty$. The equilibrium finite temperature black hole can then be written in Schwarzschild coordinates as
\begin{equation}
ds^2 = - D(r) d\tau^2 + \frac{dr^2}{D(r)}, \quad\quad\quad\quad\quad\quad \Phi(r) = a r,
\end{equation}
where $D(r) = r^2 - r_0^2$ and $a$ is the source (of dimension $-1$) for $\mathcal{O}_\phi$ in the equilibrium state. $r_0$ parameterises the temperature $T$ of the state via 
\begin{equation}
r_0=2\pi T.
\end{equation}
It will be convenient to introduce ingoing coordinates $(t,r)$ where $d\tau = dt - dr/D(r)$ such that
\begin{equation}\label{eq:EF_JT}
ds^2 = - D(r) dt^2 + 2 dt dr, \quad\quad\quad\quad\quad\quad \Phi(r) = a r.
\end{equation}

\paragraph{}We will now go through a simplified version of the steps of \cite{Blake:2026jue} to construct a hydrodynamic action describing perturbations around this equilibrium state, up to quadratic order in perturbations. The simplification is that we will work with only a single branch of the CGL contour (i.e.~a single exterior region). This will be sufficient to obtain the retarded (or advanced) two point functions of the stress tensor, which in this state are real in momentum space up to an $i\epsilon$ prescription.

\subsubsection*{Perturbations} 

\paragraph{} To proceed, we consider perturbations of the bulk fields
\begin{equation}
\delta ds^2 =  \delta g_{tt}(t,r) dt^2 + 2 \delta g_{tr}(t,r) dt dr + \delta g_{rr}(t,r) dr^2,
\end{equation}
and
\begin{equation}
\Phi(t,r) = a r + \delta \Phi(t, r), 
\end{equation}
where we decompose the metric perturbations as
\begin{equation}\label{metricansatz}
\begin{split}
\delta g_{tt}(t, r)&= -D(r) h_{tt}(t,r)+ 2\partial_t\zeta_t(t,r) - D'(r)\big(\zeta_t(t,r)+ D(r)\zeta_r(t,r)\big), \\
\delta g_{tr}(t,r) &=  D'(r)\zeta_r(t,r) + \partial_r\zeta_t(t,r) +  \partial_t\zeta_r(t,r), \\
\delta g_{rr}(t,r) &= 2\partial_r\zeta_r(t,r),
\end{split}
\end{equation}
and the dilaton perturbation as
\begin{equation}
 \delta \Phi(t, r) = a (\zeta_t(t,r)  + D(r) \zeta_r(t,r)) + \delta \phi(t,r). 
 \label{dilatonansatz} 
\end{equation}
This ansatz has a simple interpretation -- the fields $h_{tt}$ and $\delta\phi$ respectively correspond to metric and dilaton perturbations in radial gauge. The fields $\delta g_{M N}$ and $\delta \Phi$ are then obtained by acting on these radial gauge perturbations with linearised diffeomorphisms generated by $\zeta_t$ and $\zeta_r$. It will be convenient to raise indices on the diffeomorphism fields
\begin{equation}
\zeta^t=\zeta_r,\qquad\quad\quad\quad\quad \zeta^r=\zeta_t+D(r)\zeta_r,
\end{equation}
and we assume that $\zeta^M$ are regular at the future horizon.

\subsubsection*{Hydrodynamic variables}
\paragraph{} Our goal is to construct a $1$-dimensional hydrodynamic action of the form
\begin{equation}
S_{\rm hydro}[\xi_t,\delta g^{(s)}_{tt},\delta\phi^{(s)}],
\label{hydroaction}
\end{equation}
where $\xi_t(t)$ is the dynamical variable, and $\delta g^{(s)}_{tt}(t)$ and $\delta\phi^{(s)}(t)$ are linearised external sources for the stress tensor and $\mathcal{O}_\phi$ respectively. We will see that the natural hydrodynamic variable that emerges in our prescription is the relative time diffeomorphism between the horizon and
the asymptotic boundary
\begin{equation}
\xi_t(t)
= \big[-\zeta^t\big]^{\infty}_{r_0}
=
- \zeta^t(t, r \to \infty) + \zeta^{t}(t, r_0). 
\label{eq:hydro_variable_generic_JT}
\end{equation}
By integrating the last equation in \eqref{metricansatz}, the hydrodynamic variable can be equivalently expressed as a Wilson line-like field
\begin{equation}
\xi_t(t) = - \int_{r_0}^{\infty} \frac{\delta g_{rr}(t,r)}{2} dr. 
\end{equation}
\paragraph{} Now to compute the hydrodynamic action we expand $S_{\mathrm{JT}} = S_{\mathrm{EH}} + S_{\mathrm{GH}}$ in powers of $\delta g_{M N}$ and $\delta \Phi$. We will work up to quadratic order in these perturbations such that
\begin{equation}
S_{\rm{JT}}= S_0 + S_1[\delta g_{M N}, \delta \Phi] + S_2[\delta g_{M N}, \delta \Phi] + \dots,
\label{expandaction}
\end{equation}
where $S_0$ is the action evaluated on the bulk solution, $S_1$ is linear in the amplitude of the perturbations and $S_2$ is quadratic. The dots above indicate cubic and higher order terms that we will neglect. 
\subsubsection*{Equations of motion} 

\paragraph{}To obtain the hydrodynamic action~\eqref{hydroaction}, we evaluate the gravitational action \eqref{expandaction} partially on-shell. That is, by imposing some (but not all) of the bulk equations of motion for the perturbations. It is useful to first clarify the structure of the linearised equations of motion (which are precisely the equations of motion of the quadratic action $S_2[\delta g_{M N}, \delta \Phi]$). 

\paragraph{}Due to gauge invariance, the four equations of motion \eqref{metriceom} and \eqref{scalareom} only depend on the two fields $\delta\phi(t,r)$ and $h_{tt}(t,r)$ appearing in the perturbation ansatzes~\eqref{metricansatz} and~\eqref{dilatonansatz}, and not on $\zeta^M$. The four equations of motion can be arranged into two equations of motion that are second order in radial derivatives (the dynamical equations), and two that are first order in radial derivatives (the constraint equations). We take the dynamical equations to be
\begin{equation}
\delta E_{\Phi} = \partial_r^2 \Big(D(r) h_{tt}\Big) = 0, \qquad\qquad \delta E_{rr} = \partial_r^2 \delta \phi = 0.
\label{secondorder}
\end{equation}
And we take the constraint equations to be the following two linear combinations of the Einstein equations
\begin{align}
\mathrm{trace} &\equiv 2 ( \delta E_{tr} + D(r) \delta E_{rr} ) =0
\label{trace}, \\
\mathrm{energy} &\equiv -2( \delta E_{tt} + D(r) \delta E_{tr} ) - r \partial_{t}\left(\mathrm{trace}\right)  = 0.
\label{energy}
\end{align}
These have been so named because the first equation can be interpreted as ensuring that the boundary stress tensor obeys the Ward identity associated to Weyl transformations, while the latter ensures energy is covariantly conserved (in the absence of a time-dependent source for the operator $\mathcal{O}_\phi$). We will see this explicitly soon.

\paragraph{}Once the dynamical equations \eqref{secondorder} are imposed, the two constraints are radially conserved. To see this, note that
\begin{align}
\partial_r\,{\rm trace}
&=
a\delta E_\Phi
+2(r+\partial_t)\delta E_{rr} ,
\\
\partial_r\,{\rm energy}
&=
-ar\partial_t \delta E_\Phi
-2\left(r\partial_t+r_0^2\right)\partial_t\delta E_{rr} .
\end{align}
Explicitly, the constraints may be written
\begin{eqnarray}
\mathrm{trace} &=& 2  \partial_{t} \partial_{r} \delta \phi + 2 r^2 \partial_{r}(\delta \phi/r) + a \partial_{r} (D(r) h_{tt}), \nonumber \\
\mathrm{energy} &=& \partial_t\left(-2 r^2\partial_{r} ( \partial_{t} \delta \phi/r) - a r^2 \partial_{r} (D(r) h_{tt}/r) - 2 r_0^2 \partial_{r} \delta \phi\right).
\label{traceenergyexplicit}
\end{eqnarray}

\subsection*{Partially on-shell solutions}

\paragraph{} To construct the hydrodynamic action we go partially on-shell. Specifically, this means that we impose the second order equations of motion~\eqref{secondorder} as well as the trace constraint~\eqref{trace}.\footnote{We will see later that this produces a hydrodynamic action that, even off-shell, obeys the expected Ward identity for Weyl transformations.} Note that this differs from \cite{Maldacena:2016upp}, in which only the dilaton equation $E_\Phi$ is imposed in obtaining the Schwarzian action.

\paragraph{}It is straightforward to find the general solution to the second order differential equations \eqref{secondorder}
\begin{equation}\label{eq:JT_Dir_partial_solution}
D(r) h_{tt}(t,r)  =  \alpha_1(t) r + \alpha_2(t), \hspace{2.0cm} \delta \phi(t,r) = \alpha_3(t) + \alpha_4(t)r,
\end{equation}
for arbitrary functions $\alpha_i(t)$.\footnote{It is at this step where the method of \cite{Blake:2026jue} simplifies in JT gravity compared to generic theories: it is trivial to find exact solutions to the dynamical equations as there are no bulk gravitons.} For these solutions, the constraint equations~\eqref{trace} and~\eqref{energy} are radially independent as expected
\begin{equation}\label{eq:tr_JT}
\mathrm{trace} = a \alpha_1 - 2 \alpha_3 + 2 \alpha_4' = 0, \qquad \mathrm{energy} = a \alpha_2' - 2 r_0^2 \alpha_4' + 2 \alpha_3'' = 0,
\end{equation}
where primes denote derivatives with respect to the argument (in this case $t$). A general partially on-shell solution is then specified by four integration constants $\alpha_i(t)$ for $i=1,2,3,4$. We also impose the trace constraint equation which can be thought of as fixing $\alpha_3$ in terms of $\alpha_1$ and $\alpha_4$. Thus, the partially on-shell solution can be parameterised by three independent functions $\alpha_1(t)$, $\alpha_2(t)$ and $\alpha_4(t)$. In terms of these integration constants, the energy constraint becomes
\begin{equation}
\mathrm{energy} = a \alpha_1'' + a \alpha_2' - 2 r_0^2 \alpha_4' + 2\alpha_4''' = 0. 
\label{eq:bulk_energy}
\end{equation}
As we discussed previously, we do not impose this when deriving the hydrodynamic action. It will ultimately arise as the equation of motion obtained by varying the hydrodynamic action \eqref{hydroaction} with respect to $\xi_t$.

\paragraph{} It is helpful to note that an alternative way to think about the above prescription is that we are imposing the dynamical equations of motion~\eqref{secondorder} together with the Einstein equation $\delta E_{tr} = 0$. The remaining Einstein equation $\delta E_{tt} = 0$ is then left off-shell, and can be seen from~\eqref{energy} to be equivalent (up to a factor of $-2$) to the energy constraint \eqref{eq:bulk_energy}. 

\subsubsection*{Asymptotic boundary conditions} 

\paragraph{} We now wish to fix the integration constants $\alpha_1(t)$, $\alpha_2(t)$ and $\alpha_4(t)$ in our partially on-shell solutions by imposing boundary conditions at the asymptotic boundary and the horizon. Starting with the boundary, we first expand the diffeomorphism fields for large $r$ as
\begin{align}
\zeta_M(t,r) &= \zeta_M^{(2)}(t) r^2+\zeta_M^{(1)}(t) r+\zeta_M^{(0)}(t) +\zeta_M^{(-1)}(t) r^{-1}+\ldots.
\label{diffeouvads2}
\end{align}
We then restrict the coefficients in this expansion in such a way that upon switching to Schwarzschild coordinates the partially on-shell solutions are in Fefferman-Graham gauge. This will allow us to easily identify the external sources using the usual holographic dictionary.

\paragraph{} In particular, letting $\delta G_{MN}$ and $\delta\tilde{\Phi}$ denote the Fourier modes of the metric and dilaton perturbations in Schwarzschild coordinates we have 
\begin{equation}
\begin{split}
\delta G_{\tau\tau} &= e^{- i \omega r_*} \delta g_{tt}, \quad\quad\quad\quad\quad\quad\quad\quad\quad\quad\quad\quad \delta G_{\tau r}= e^{- i \omega r_*} \bigg(\delta g_{tr} + \frac{\delta g_{tt}}{D(r)}\bigg), \\
\delta G_{rr} &= e^{- i \omega r_*} \bigg(\delta g_{rr}+ \frac{2 \delta g_{tr}}{D(r)} + \frac{\delta g_{tt}}{D(r)^2}\bigg),\,\quad\quad\;\; \delta\tilde{\Phi}=e^{-i\omega r_*}\delta\Phi,
\end{split}
\end{equation}
with $r_*=\int^r_{\infty} d\rho/D(\rho)$ being the tortoise coordinate. For our purposes it is sufficient to impose Fefferman-Graham gauge by demanding $\delta G_{\tau r} \sim {\cal O}(1/r)$ and $\delta G_{rr}\sim {\cal O}(1/r^3)$, which requires\footnote{This gauge can be imposed at successively higher orders in the large $r$ expansion by fixing further coefficients in the expansion~\eqref{diffeouvads2} in terms of $\zeta_r^{(-1)}$ and $\zeta_t^{(2)}$. }
\begin{equation}
\zeta_r^{(2)} = \zeta_r^{(1)} = 0, \hspace{1.5cm} \zeta_r^{(0)} = - \zeta_{t}^{(2)}, \hspace{1.5cm} \zeta_t^{(1)} = \partial_{t} \zeta_t^{(2)} .
\end{equation}
For our partially on-shell solutions we then find that
\begin{equation}
\begin{aligned}
\delta G_{\tau\tau}(r\rightarrow\infty)&\,\rightarrow -2\zeta_r^{(-1)}\,r^2+O(r),\\
\delta \tilde{\Phi}(r\rightarrow\infty)\,&\rightarrow 
\left(
\alpha_4
+a\zeta_r^{(-1)}
+a\partial_t\zeta_t^{(2)}
\right)r
+O(r^0).
    \end{aligned}
\end{equation}
From these we use the usual holographic dictionary to identify the linearised perturbations of the field theory metric $\delta g_{tt}^{(s)}$ and scalar source $\delta\phi^{(s)}$ as\footnote{If we interpret hydrodynamics strictly as a universal theory for the stress tensor then we should set $\delta\phi^{(s)}(t)$ to zero, which can be done at any stage. We allow it to be non-trivial for completeness, ultimately yielding the two-point functions of $\mathcal{O}_{\phi}$ as well as of $T^{tt}$.}
\begin{equation}
\begin{aligned}
\label{UVbc}
  \delta g_{tt}^{(s)}(t)&\,=-2\zeta_r^{(-1)}(t),\\
  \delta\phi^{(s)}(t)&\,=\alpha_4(t)-\frac{a}{2}\delta g_{tt}^{(s)}(t)+a\partial_t\zeta_t^{(2)}(t).
\end{aligned}
\end{equation}
We use these equations to fix the functions $\zeta_r^{(-1)}$ and $\alpha_4$.

\subsubsection*{Horizon boundary conditions} 

\paragraph{}To obtain a hydrodynamic action of the form \eqref{hydroaction} we need to fix the remaining integration constants $\alpha_1, \alpha_2$. To do this, we impose boundary conditions at the horizon. As mentioned in Section \ref{sec:hor_sy}, there is more than one way of doing this that produces a consistent hydrodynamic action (i.e.~an action of the form \eqref{hydroaction} that produces the correct two-point functions). In the main text we will consider mixed boundary conditions that are consistent with the variational principle when a Gibbons-Hawking term is not added at the horizon. These are analogous to the boundary conditions studied in \cite{Blake:2026jue}, and in JT gravity we will show they produce the Schwarzian action with the horizon symmetries of \cite{Knysh:2024asf}. In Appendix \ref{app:Dir} we will consider adding a Gibbons-Hawking term near the horizon and then imposing Dirichlet boundary conditions, which yields a different form of the hydrodynamic action with different horizon symmetries. 

\paragraph{}The horizon boundary conditions that we impose are
\begin{equation}
\delta g_{tt}({r_0)}=0,\qquad
\delta\kappa(r_0)\equiv
\partial_r\delta g_{tt}(r_0)
-2\partial_t\delta g_{tr}(r_0)
+2r_0\delta g_{tr}(r_0)
=0,
\label{eq:generic-gauge-horizon-bcs}
\end{equation}
where $\delta\kappa$ can be interpreted as the change in surface gravity.\footnote{These horizon boundary conditions, and the remainder of the calculation of the hydrodynamic action, become simpler when working in radial gauge near the horizon, as we show in Appendices \ref{app:radial-gauge-actions} and \ref{app:radial-gauge-horizon-symmetries}.} The motivation for these boundary conditions is straightforward and shown in Appendix~\ref{gen_bcs_JT-new}: they respect the variational principle of the quadratic Einstein-Hilbert action at the horizon. This ensures that varying the hydrodynamic action with respect to  $\xi_t$ will yield the energy constraint equation, and thus upon integrating out the hydrodynamic field we are studying solutions to the full set of gravitational equations of motion. 

\paragraph{} Imposing \eqref{eq:generic-gauge-horizon-bcs} fixes the remaining integration constants $\alpha_1(t)$ and $\alpha_2(t)$ to be
\begin{equation}
\begin{split}
\alpha_1(t) &= - 2 \zeta^{r}(t,r_0)- 2 r_0 \partial_{t} \zeta^{t}(t,r_0) - 2 \partial_{t}^2 \zeta^{t}(t, r_0), \nonumber \\
\alpha_2(t) &= 2 \partial_{t} \zeta^{r}(t,r_0) + 2 r_0^2 \partial_{t} \zeta^{t}(t,r_0) + 2 r_0 \partial_{t}^2 \zeta^{t}(t,r_0).
\end{split}\label{eq:gen_alp12}
\end{equation}
With these identifications the energy constraint~\eqref{eq:bulk_energy} -- which we have not yet imposed -- becomes
\begin{equation}
\begin{split}
\mathrm{energy}=
(\partial_t^2 - r_0^2)\partial_t
\left(
2\delta\phi^{(s)}
+a\left(\delta g^{(s)}_{tt}-2\partial_t\xi_t\right)
\right)= 0, 
\end{split}
\label{energy_explicit}
\end{equation}
where $\xi_t(t) = [-\zeta^{t}]^{\infty}_{r_0} = \zeta_t^{(2)}(t) + \zeta^{t}(r_0)(t)$ is the relative diffeomorphism we advertised earlier in equation \eqref{eq:hydro_variable_generic_JT}.

\subsubsection*{Linear hydrodynamic action}

\paragraph{} We now proceed to compute the hydrodynamic action by evaluating the gravitational action for the partially on-shell solutions. We start with the linearised action $S_1[\delta g_{MN}, \delta \Phi]$ obtained by expanding $S_{\mathrm{JT}}$ to linear order in perturbations. This yields
\begin{equation}\label{eq:linJT}
S_{\mathrm{hydro}}^{(1)}[\xi_t, \delta g_{tt}^{(s)}, \delta \phi^{(s)}]  = \int dt \left(r_0^2 \delta \phi^{(s)} + \frac{a r_0^2}{2} \delta g_{tt}^{(s)}  \right),
\end{equation}
where here (and throughout) we integrate by parts and neglect terms that are total $t$ derivatives.

\subsubsection*{Quadratic hydrodynamic action} 

\paragraph{} We now wish to evaluate the quadratic action $S_2[\delta g_{MN}, \delta \Phi]$ on our partially on-shell solutions. To do this, we note that the quadratic piece of the Einstein-Hilbert action can be expressed as 
\begin{equation*}
S_{\mathrm{EH}} = S_{\mathrm{bulk}} +  S_{\mathrm{boundary}}(\infty) - S_{\mathrm{boundary}}(r_0),
\end{equation*}
where the bulk contribution is given by 
\begin{equation*}
 S_{\mathrm{bulk}}  = \frac{1}{2} \int d^2 x\left(\delta E_{\Phi} \delta \Phi + \delta E_{M N} \delta g^{M N}\right),
\end{equation*}
with $\delta E_{\Phi}$ and $\delta E_{M N}$ the linearised equations of motion, and we will return to the boundary term shortly.

\paragraph{}For our partially on-shell solutions, we imposed $\delta E_\Phi=\delta E_{tr}=\delta E_{rr}=0$. Therefore the bulk contribution reduces to
\begin{equation}\label{eq:ene_mom}
S_{\mathrm{bulk}} = -\frac{1}{2} \int d^2x \delta E_{tt} \delta g_{rr} = \frac{1}{2} \int d^2x \partial_{r}\left(\mathrm{energy} \times \zeta^t\right)= -\frac{1}{2} \int dt \left(\xi_t\times \mathrm{energy}\right),
\end{equation} 
where in the second equality we used the fact that the energy constraint is radially conserved for partially on-shell solutions, and in the third equality we used the definition \eqref{eq:hydro_variable_generic_JT} of the hydrodynamic variable $\xi_t$ as a relative diffeomorphism. From the explicit form of the energy equation in~\eqref{energy_explicit}, we see that this part of the action depends only on the hydrodynamic field $\xi_t$ and the sources $\delta g_{tt}^{(s)}$ and $\delta\phi^{(s)}$.

\paragraph{}To compute the full action we have to also consider the contributions from $S_{\mathrm{boundary}}$ at the asymptotic boundary and the horizon, and from the Gibbons-Hawking term at the asymptotic boundary. Explicitly, the boundary term is
\begin{equation}
\begin{split}
 S_{\mathrm{boundary}} =\int dt&\Biggl( - \frac{3}{4} a r^2 D(r)^2 \delta g_{rr}^2 - 2 a r^2 \delta g_{tr}^2 + \frac{a}{2} \delta g_{tr} \delta g_{tt} - a r \delta g_{tr} \delta g_{tt,r} \\
 &-\frac{1}{2} \delta g_{tt} \delta \Phi_{,r} + \frac{1}{2} \delta g_{tt,r} \delta \Phi + \frac{r}{2} D(r) \delta \Phi \delta g_{rr} - 3 a r^2 D(r) \delta g_{rr} \delta g_{tr} \\
 &+ a rD(r) \delta g_{rr} \delta g_{tr,t} - \delta g_{tr,t} \delta \Phi + r \delta \Phi \delta g_{tr} - \frac{1}{2} a r \delta g_{tt} \delta g_{rr,t} \\
 &- \frac{a}{4} (3 r^2 + r_0^2) \delta g_{tt} \delta g_{rr} - \frac{a r}{2}D(r) \delta g_{rr} \delta g_{tt,r}\Biggr), 
\end{split}\label{boundary1}
\end{equation}
where commas denote partial derivatives with respect to the subsequent coordinate. This vanishes at the horizon (up to a total derivative $-ar_0\partial_t(\left.\delta g_{tr}^2\right|_{r_0})$) under the horizon boundary conditions \eqref{eq:generic-gauge-horizon-bcs}. We therefore only get additional contributions to the action from the asymptotic boundary. Including these terms we find the total quadratic hydrodynamic action
\begin{equation}
\begin{aligned}
 S_{\mathrm{hydro}}^{(2)} = \int dt \Biggl(&\,a (\xi_t''^2 + r_0^2 \xi_t'^2) + 2 \bigg(\delta \phi^{(s)} + a \frac{\delta g_{tt}^{(s)}}{2}\bigg)(\xi_t''' - r_0^2 \xi_t') \\
 &\,+\frac{3 a r_0^2}{8} \delta {g_{tt}^{(s)}}^2 - \frac{a}{4} \delta {g_{tt}^{(s)}}'' \delta g_{tt}^{(s)} +  \frac{r_0^2}{2} \delta \phi^{(s)} \delta g_{tt}^{(s)} -  \delta \phi^{(s)} \delta {g_{tt}^{(s)}}''\Biggr).
 \label{hydroJT}
 \end{aligned}
\end{equation}
This is the central result of this Section.

\paragraph{}In the absence of external sources, the hydrodynamic action reduces to
\begin{equation}
 S_{\mathrm{hydro}}[\xi_t] = a\int dt \, \left(\xi_t''^2 + r_0^2 \xi_t'^2\right), 
 \label{hydroJTsourcefree}
\end{equation}
which is precisely the quadratic Schwarzian action for JT gravity, in Lorentzian signature, obtained in \cite{Maldacena:2016upp}. Note that the quadratic in $\xi_t$ terms arise from $S_{\mathrm{bulk}}$ (that is, the boundary terms all vanish in the absence of external sources). While the result is familiar the interpretation of $\xi_t$ is different, as is the method which in principle can be straightforwardly generalised to higher dimensions \cite{Blake:2026jue}. Note that the horizon boundary conditions \eqref{eq:generic-gauge-horizon-bcs} are crucial for obtaining the above form of the action. In Appendix \ref{app:Dir} we consider imposing Dirichlet boundary conditions at the horizon (similar to~\cite{deBoer:2015ija}), and find an action that is not given by the Schwarzian (but nevertheless gives the same two-point functions). 

\paragraph{}In the approach of  \cite{Maldacena:2016upp}, the Euclidean Schwarzian action was obtained from Euclidean JT gravity, where there is no horizon (and therefore no Gibbons-Hawking term there). Working with the natural boundary conditions \eqref{eq:generic-gauge-horizon-bcs} in the absence of a Gibbons-Hawking term at the horizon is the Lorentzian analogue of this. For this reason, it makes sense that it is these boundary conditions (alongside the absence of a horizon Gibbons-Hawking term) that produce the Lorentzian Schwarzian action.

\subsection*{Compact forms of the quadratic action}

\paragraph{} In order to emphasise the connection of the Schwarzian to hydrodynamic actions, and to make some of its symmetries in the presences of sources manifest, we will now write the action in \eqref{hydroJT} in a more compact form using pullbacks of the sources. In particular, in arbitrary dimensions, let us define the fields ${\mathcal B}_{\mu \nu}$ to be the perturbations of the induced boundary metric under (boundary) diffeomorphisms 
\begin{equation}\label{eq:curly_B}
{\mathcal B}_{\mu \nu}(x) = \frac{\partial y^{\alpha}}{\partial x^{\mu} } \frac{\partial y^{\beta}}{\partial x^{\nu}}  g^{(s)}_{\alpha \beta}(y(x)) - \eta_{\mu \nu},
\end{equation}
where $y^{\mu}= x^{\mu} - \xi^{\mu}$ and $g_{\mu \nu}^{(s)} = \eta_{\mu \nu} + \delta g_{\mu \nu}^{(s)}$. Now, specialising to ($0+1$)-dimensions and denoting the temporal component: $y^t$ as $y(t)$, we get
\begin{equation}
y(t)=t-\xi^t(t)=t+\xi_t(t), \qquad\qquad g_{tt}^{(s)}(t)=-1+\delta g_{tt}^{(s)}(t),
\end{equation}
where we raise and lower indices on the hydrodynamic field with the Minkowski metric. Now under the pullback action: $t\mapsto y(t)$,
\begin{equation}
(ds^2)_{\rm pullback}
=
g_{tt}^{(s)}(y(t))\,dy^2
=
\left(\frac{dy}{dt}\right)^2 g_{tt}^{(s)}(y(t))\,dt^2,
\end{equation}
and so at quadratic order in perturbations $\mathcal{B}(t)$ above is
\begin{equation}\label{eq:B_lin_quad}
\mathcal B = \delta g_{tt}^{(s)} - 2\partial_t\xi_t + \xi_t\partial_t\delta g_{tt}^{(s)} + 2\delta g_{tt}^{(s)}\partial_t\xi_t - (\partial_t\xi_t)^2,
\end{equation}
in the (0+1)-dimensional case. Similarly, for the perturbations of the dilaton source we define
\begin{equation}\label{eq:curly_phi}
\varphi(t)=\delta\phi^{(s)}(y(t)) = \delta\phi^{(s)} + \xi_t\partial_t\delta\phi^{(s)}.
\end{equation}
Then the hydrodynamic action to quadratic order can be written compactly as
\begin{equation}
\begin{aligned}
\label{B_cpt}
S_{\mathrm{hydro}}
=
\int dt\left(
r_0^2\varphi+\frac{ar_0^2}{2}\mathcal B+\frac{3ar_0^2}{8}\mathcal B^2
-\frac{a}{4}\mathcal B\,\partial_t^2\mathcal B
+\frac{r_0^2}{2}\varphi\mathcal B
-\varphi\,\partial_t^2\mathcal B
\right).
\end{aligned}
\end{equation}
As we will soon show explicitly, this form of the hydrodynamic action is manifestly diffeomorphism invariant and ensures that the hydrodynamic equation of motion is the local conservation of energy.

\paragraph{}The action can be written even more compactly by first defining the fields
\begin{equation}
\mathcal{G}_{tt}(t)\equiv  \left(\frac{dy}{dt}\right)^2  g^{(s)}_{tt}(y(t)),\quad\quad\quad\quad\quad\quad \Psi(t)\equiv \phi^{(s)}(y(t)),
\end{equation}
and the ratio
\begin{equation}
\label{eq:sigmatdefn}
    \sigma(t)\equiv\frac{\Psi(t)}{\sqrt{-\mathcal{G}_{tt}(t)}},
\end{equation}
which is invariant under the Weyl transformations $\phi^{(s)}(y(t))\rightarrow e^{-\Omega(y(t))}\phi^{(s)}(y(t))$ and $g^{(s)}_{tt}(y(t))\rightarrow e^{-2\Omega(y(t))}g^{(s)}_{tt}(y(t))$. Explicitly, to quadratic order in perturbations
\begin{equation}
\begin{aligned}
\label{eq:omegadefn}
    \sigma(t)=&\,a+\left(\delta\phi^{(s)}(t)+\frac{a}{2}\delta g_{tt}^{(s)}(t)-a\partial_t\xi_t(t)\right)\left(1-\partial_t\xi_t(t)\right)+\frac{a}{2}\xi_t(t)\partial_t\delta g_{tt}^{(s)}(t)\\
    &\,+\xi_t(t)\partial_t\delta\phi^{(s)}(t)+\frac{3a}{8}{\delta g_{tt}^{(s)}(t)}^2+\frac{1}{2}\delta\phi^{(s)}(t)\delta g_{tt}^{(s)}(t),\\
    \Psi(t)=&\,a+\delta\phi^{(s)}(t)+\xi_t(t)\partial_t\delta\phi^{(s)}(t).
    \end{aligned}
\end{equation}

\paragraph{}In terms of these fields, our hydrodynamic action can be written up to quadratic order in perturbations as\footnote{The equilibrium (zeroeth order in perturbation) part of this action $ar_0^2\int dt$ gives the correct value of the thermal partition function upon analytic continuation to Euclidean signature.}
\begin{equation}
    \label{quadomegaaction}
    S_{\mathrm{hydro}}=\int dt\left(r_0^2\sigma-\frac{1}{a}\sigma\partial_t^2\sigma+\frac{1}{a}\Psi\partial_t^2\Psi\right).
\end{equation}
This form  isolates the Weyl anomaly of the hydrodynamic theory in the last term, as we will explain in detail shortly. The first term can be understood as simply
\begin{equation}
\label{eq:hydrostatic-act}
   \int dt\sqrt{-g^{(s)}}P[g^{(s)},\phi^{(s)}],\quad\quad P[g^{(s)},\phi^{(s)}]=4\pi^2(-g_{\mu\nu}^{(s)}\beta^\mu\beta^\nu)^{-1}\phi^{(s)}, 
\end{equation}
where the metric $g^{(s)}(t)$ and scalar source $\phi^{(s)}(t)$ are replaced by $\mathcal{G}(t)$ and $\Psi(t)$ respectively. In equation 
\eqref{eq:hydrostatic-act}, $P$ is a covariant representation of the equilibrium pressure $P=r_0^2\phi^{(s)}$ where the thermal vector $\beta^\mu\equiv\frac{u^\mu}{T}$ with $g_{\mu\nu}^{(s)}u^\mu u^\nu=-1$. Relations between hydrodynamic Lagrangians and thermodynamic potentials are expected in general for non-dissipative fluids \cite{Dubovsky:2011sj}.

\subsection*{Constitutive relations, equations of motion and two-point functions}

\paragraph{}For now, we will proceed with the explicit action given by the sum of \eqref{eq:linJT} and \eqref{hydroJT}. From this we will derive expressions for the stress tensor and $\mathcal{O}_\phi$ in terms of the hydrodynamic field and show that, when this field is placed on-shell, their two-point functions coincide with the known results.

\paragraph{}We define the (one-dimensional) stress tensor and scalar expectation values through 
\begin{equation}\label{eq:vevs_T_phi}
T^{tt}
=
\frac{2}{\sqrt{-g^{(s)}}}
\frac{\delta S_{\rm hydro}}{\delta(\delta  g^{(s)}_{tt})},
\qquad\qquad\qquad
{\cal O}_\phi
=
\frac{1}{\sqrt{-g^{(s)}}}
\frac{\delta S_{\rm hydro}}{\delta (\delta\phi^{(s)})} .
\end{equation}
From the linear action \eqref{eq:linJT} we get their values in equilibrium (i.e.~in the absence of sources $\delta g_{tt}^{(s)}$ and $\delta\phi^{(s)}$)
\begin{equation}\label{eq:eqm_val}
T^{tt}_{\rm eq}=a r_0^2,
\qquad\qquad\qquad
\mathcal O_\phi^{\rm eq}=r_0^2,
\end{equation}
and from the quadratic action we obtain the following constitutive relations in terms of the hydrodynamic field and sources
\begin{equation}
\begin{split}
\delta T^{tt} 
&=a\left(\partial_t^2-r_0^2\right)\left(2\partial_t\xi_t-\delta g_{tt}^{(s)}\right)+ar_0^2\delta g_{tt}^{(s)}+r_0^2\delta\phi^{(s)}-2\partial_t^2\delta\phi^{(s)}, \\
\delta\mathcal O_\phi 
&=\left(\partial_t^2-r_0^2\right)\left(2\partial_t\xi_t-\delta g_{tt}^{(s)}\right). 
\end{split}\label{eq:off_var}
\end{equation}
These satisfy the following relation identically (i.e.~even when $\xi_t$ is off-shell)
\begin{equation}\label{eq:Ward_id_JT}
\delta T^{tt}-a\delta {\cal O}_\phi = ar_0^2 \delta g_{tt}^{(s)} +r_0^2\delta\phi^{(s)} -2\partial_t^2\delta\phi^{(s)}.
\end{equation}
This is the linearised version of the Ward identity arising from Weyl transformations $g_{tt}^{(s)}T^{tt}+\phi^{(s)}\mathcal{O}_\phi=-2\nabla^2\phi^{(s)}$ where the Laplacian is that of $g^{(s)}$.\footnote{This is consistent with equation (3.16) of \cite{Cvetic:2016eiv} after noting that their $\mathcal{O}_\psi=\phi^{(s)}\mathcal{O}_\phi$.} The term on the right hand side is a (matter) conformal anomaly.

\paragraph{}To progress we must put the hydrodynamic field on-shell. Its equation of motion is
\begin{equation}
\begin{aligned}
\label{eq:hydro_eom}
0=\frac{\delta S_{\rm hydro}}{\delta \xi_t} &\,=-(\partial_{t}^2 - r_0^2)\partial_t\left(2\delta\phi^{(s)}+a\left(\delta g_{tt}^{(s)}-2\partial_t\xi_t\right)\right) \\
&\,= \partial_t\delta T^{tt} -ar_0^2\,\partial_t\delta g^{(s)}_{tt} +r_0^2\,\partial_t\delta\phi^{(s)}.
\end{aligned}
\end{equation}
By comparing the first line to~\eqref{energy_explicit}, we see that the hydrodynamic equation of motion is (up to a minus sign) the remaining Einstein constraint equation that we did not impose when we put the fields partially on-shell. In other words, imposing the hydrodynamic equation of motion leaves us with the usual holographic generating functional we would obtain by solving all of the classical bulk equations. The second line is the linearisation of the Ward identity $\nabla_t T^{tt}=\mathcal{O}_\phi\nabla^t\phi^{(s)}$: ~the hydrodynamic equation of motion enforces local conservation of the stress tensor. At generic frequencies $\omega\ne0,\pm ir_0$ the solution to the hydrodynamic equation is
\begin{equation}
\label{eq:freq_2D}
\xi_t(\omega)=\frac{i}{2 a \omega}\left(a\,\delta g_{tt}^{(s)}(\omega)+2\delta\phi^{(s)}(\omega)\right).
\end{equation}\vspace{0.1mm}

\paragraph{}Finally, we evaluate the response functions. On-shell the constitutive relations become
\begin{equation}
\delta T^{tt}_{\rm{on}}
=
ar_0^2\,\delta g^{(s)}_{tt}
-r_0^2\,\delta\phi^{(s)},
\qquad\qquad
\delta{\cal O}^{\rm on}_\phi
=
\frac{2}{a}\left(\partial_t^2-r_0^2\right)\delta\phi^{(s)} .
\end{equation}
Defining
\begin{equation}
\begin{aligned}
\label{eq:ads22ptfndefn}
    G_{T^{tt}T^{tt}}(\omega)&\,=2\frac{\delta \left(\sqrt{-g^{(s)}}T^{tt}\right)}{\delta g_{tt}^{(s)}},\quad\quad\quad G_{T^{tt}\mathcal{O}_\phi}(\omega)=\frac{\delta \left(\sqrt{-g^{(s)}}T^{tt}\right)}{\delta \phi^{(s)}},\\
        G_{\mathcal{O}_\phi T^{tt}}(\omega)&\,=2\frac{\delta \left(\sqrt{-g^{(s)}}\mathcal{O}_\phi\right)}{\delta g_{tt}^{(s)}},\quad\quad\quad G_{\mathcal{O}_\phi \mathcal{O}_\phi}(\omega)=\frac{\delta \left(\sqrt{-g^{(s)}}\mathcal{O}_\phi\right)}{\delta \phi^{(s)}},
    \end{aligned}
\end{equation}
we obtain the response functions
  \begin{equation}\label{eq:const_2D}
G_{T^{tt} T^{tt}}(\omega)=a r_0^2,
\qquad
G_{T^{tt} \mathcal{O}_{\phi}}(\omega)=G_{\mathcal{O}_\phi T^{tt}}(\omega)=-r_0^2,
\qquad
G_{\mathcal{O}_{\phi} \mathcal{O}_{\phi}}(\omega)=-\frac{2}{a}(\omega^2+r_0^2).
\end{equation}  
Equivalently, these can be extracted from the on-shell hydrodynamic action via
\begin{equation}
\begin{aligned}
S_{\text{hydro,on-shell}} = \int \frac{d\omega}{2 \pi} \Biggl(&\, \frac{1}{8} G_{T^{tt} T^{tt}}(\omega) \delta g_{tt}^{(s)}(-\omega)\delta g_{tt}^{(s)}(\omega)  + \frac{1}{4} G_{\mathcal{O}_\phi  T^{tt}}(\omega) \delta\phi^{(s)}(-\omega) \delta g_{tt}^{(s)}(\omega) \\
&\,+ \frac{1}{4} G_{  T^{tt}\mathcal{O}_\phi}(\omega)\delta g_{tt}^{(s)}(-\omega)\delta\phi^{(s)}(\omega)+ \frac{1}{2} G_{\mathcal{O}_\phi \mathcal{O}_\phi}(\omega)\delta\phi^{(s)}(-\omega) \delta\phi^{(s)}(\omega)  \Biggr).
\end{aligned}
\end{equation}

\paragraph{}For generic frequencies, the expressions \eqref{eq:const_2D} are the retarded (or advanced) two-point functions. By construction they are equivalent to the expressions obtained by replacing $\omega_E\rightarrow\pm i\omega$ in the frequency space Euclidean correlators defined by \eqref{eq:ads22ptfndefn}. Note also that the two-point functions follow from the definitions 
\eqref{eq:ads22ptfndefn} provided only the Weyl and diffeomorphism Ward identities \eqref{eq:Ward_id_JT} and \eqref{eq:hydro_eom} are satisfied. In this sense, the results \eqref{eq:const_2D} are correct as they are fixed entirely by field theory symmetries.\footnote{This lack of true dynamics is the field theory analogue of the fact that there are no propagating degrees of freedom in the gravitational theory.} The two-point functions are contact terms after a Fourier transform to real time. Note that different definitions of the Euclidean and/or retarded two-point functions are sometimes used, which differ from those here by contact terms.

\paragraph{}The symmetric correlator can be obtained by treating the $\omega=0$ mode more carefully. Specifically, the pole of \eqref{eq:freq_2D} should be shifted by $\pm i\epsilon$, depending on whether one is interested in the retarded or advanced correlator. This will generate a small imaginary part for each of these frequency space two-point functions and using the fluctuation-dissipation relation to construct the symmetric correlator from these will give a non-zero contact term. Note that if we had used the full two-sided prescription of \cite{Blake:2026jue} with two independent copies of the boundary conditions above, the fields on each branch would decouple, yielding a Schwinger-Keldysh action with only mixed $-+$ type terms (in the notation of that paper) that produces the retarded Green's functions given by \eqref{eq:const_2D}. To obtain the $i\epsilon$ regularisation at the level of the action (and therefore $--$ type terms) would seem to require an $i\epsilon$ prescription for the $\omega=0$ modes of the partially on-shell solutions, and it would be worthwhile to clarify this.

\subsection*{Ultraviolet symmetries}

\paragraph{}Having obtained the correlators explicitly, let's now go back and analyse the symmetries of the action. There are two types -- ultraviolet symmetries and horizon symmetries -- that are inherited from diffeomorphisms in the gravitational description.

\paragraph{}We start with the ultraviolet symmetries, which enforce the fact that the equations of motion of our hydrodynamic action are the Ward identities for the stress tensor. In general, bulk diffeomorphisms that preserve Fefferman-Graham gauge near the conformal boundary, and vanish near the horizon, act on both the hydrodynamic variable and the boundary sources (as can be seen from their definitions \eqref{eq:hydro_variable_generic_JT} and \eqref{UVbc}). 

\paragraph{}Consider first a temporal diffeomorphism, which is simply a shift $\xi_t(t)\rightarrow\xi_t(t)+\epsilon_t(t+\xi_t(t))$ along with a compensating transformation of $g_{tt}^{(s)}$ and $\phi^{(s)}$ such that the fields $\mathcal{B}(t)$ and $\varphi(t)$ are invariant. The hydrodynamic action is invariant under such a temporal diffeomorphism, as is manifest from equation \eqref{B_cpt}. In general this is not a symmetry of the action in the conventional sense, as it acts on the sources as well as the field. However, it ensures that the hydrodynamic equation of motion is covariant conservation of the stress tensor. Explicitly, the transformation (to linear order in both $\epsilon_t$ and perturbations of the fields and sources) acts as
\begin{equation}
\label{eq:UV_time_diffeo}
\delta_\epsilon\xi_t=\epsilon_t+\xi_t\partial_t\epsilon_t,\quad\delta_\epsilon\delta g_{tt}^{(s)}=2\partial_t\epsilon_t-2\delta g_{tt}^{(s)}\partial_t\epsilon_t-\epsilon_t\partial_t\delta g_{tt}^{(s)},\quad\delta_\epsilon\delta\phi^{(s)}=-\epsilon_t\partial_t\delta\phi^{(s)},
\end{equation}
from which it is straightforward to explicitly check that \eqref{eq:hydro_eom} follows from invariance of the hydrodynamic action $\delta_\epsilon S_{\text{hydro}}=0$.

\paragraph{}We can also consider an asymptotic radial diffeomorphism $\chi^r=-r\Omega(t)+\ldots$, with subleading terms in $\chi^M$ to preserve Fefferman-Graham gauge. These act on the sources as Weyl transformations
\begin{equation}
\label{eq:2dsourceweyltransforms}
    \delta_\Omega \delta g_{tt}^{(s)}=2\Omega-2\Omega\delta g_{tt}^{(s)},\quad\quad\delta_\Omega\delta\phi^{(s)}=-a\Omega-\Omega\delta\phi^{(s)},
\end{equation}
to linear order in both $\Omega(t)$ and perturbations of the fields and sources. The field $\sigma(t)$ defined in \eqref{eq:omegadefn} is Weyl-invariant under the transformations in \eqref{eq:2dsourceweyltransforms} with $\xi_t$ held fixed, which makes it easy to see from equation \eqref{quadomegaaction} that the hydrodynamic action transforms anomalously $\delta S_{\mathrm{hydro}}/\delta\Omega=-2\partial_t^2\delta\phi^{(s)}$. Expressing the left hand-side in terms of the variation with respect to the sources, we see that this anomalous Weyl transformation ensures the Ward identity \eqref{eq:Ward_id_JT} is satisfied by the hydrodynamic action off-shell.

\subsection*{Horizon symmetries}

\paragraph{}Now we turn to the horizon symmetries, which are analogues of the asymptotic symmetries but for bulk diffeomorphisms that vanish asymptotically and are non-trivial near the horizon. Parameterising such an infinitesimal horizon diffeomorphism as $\chi=f(t)\partial_t+Z(t)\partial_r$, it acts on the metric near the horizon as $\delta g_{MN} \to \delta g_{MN} + \nabla_{M} \chi_N + \nabla_{N} \chi_{M}$. In our hydrodynamic action, these shift the horizon diffeomorphism fields $\delta_H\zeta^t(t,r_0)=f(t)$ and $\delta_H\zeta^r(t,r_0)=Z(t)$ and hence the hydrodynamic field $\delta_H\xi_t=f(t)$ while leaving the sources $\delta g_{tt}^{(s)}$ and $\delta\phi^{(s)}$ invariant.\footnote{More precisely we mean that the sources are invariant including at linear order in perturbations, while the horizon diffeomorphism fields transform in the stated way to zeroeth order. We do not need to consider first order corrections to the horizon diffeomorphisms as they appear only at quadratic order in the action (up to a total derivative).} For the quantities appearing in our horizon boundary conditions \eqref{eq:generic-gauge-horizon-bcs}, it acts explicitly as
\begin{equation}
\delta_H(\delta g_{tt}(t, r_0)) = - 2 r_0 Z + 2 \partial_{t} Z, \hspace{2.0cm}   \delta_H(\delta \kappa(t, r_0)) = -2 \partial_{t}^2 f - 2 r_0 \partial_{t}  f - 2 Z,
  \label{eq:JT_horizon_symmetry_conditions}
\end{equation}
to zeroeth order in perturbations. 

\paragraph{}The subset of these horizon diffeomorphisms that leave our horizon boundary conditions \eqref{eq:generic-gauge-horizon-bcs} unchanged are easily found to be
\begin{equation}
\label{eq:JThorsymsol}
f(t)=f_0+f_- e^{-r_0 t}+f_+ e^{r_0 t},\quad\quad\quad\quad\quad\quad Z(t) = - 2 r_0^2 f_+ e^{r_0 t}.
\end{equation}
These are the horizon symmetries for these boundary conditions. At zeroeth order in perturbations they act on the hydrodynamic action by transforming only the hydrodynamic field
\begin{equation}
\label{eq:ads2-xit-transform}
    \delta_H\xi_t=f_0+f_-e^{-r_0t}+f_+e^{r_0t}.
\end{equation}
Since $\delta_H\mathcal{B}=-2\partial_tf$, the exponential modes are distinct from the ultraviolet diffeomorphisms considered earlier (under which $\mathcal{B}$ was invariant).

\paragraph{}The horizon symmetry modes
\begin{equation}
    \xi_t(t)=f_0+f_- e^{-r_0 t}+f_+ e^{r_0 t},
\end{equation}
are a redundancy of our parameterisation. One immediately sees from the explicit form \eqref{eq:off_var} that the constitutive relations  are invariant under adding such modes (as is the hydrodynamic equation of motion \eqref{eq:hydro_eom}). In other words, the horizon symmetry modes carry no energy density.\footnote{In the absence of external sources, the kernel of the hydrodynamic equation of motion \eqref{eq:hydro_eom} contains one additional linear mode $\xi_t=f_1t$ that does carry energy.} However, the exponentially growing mode is important for understanding the quantum chaotic properties of the state \cite{Maldacena:2016hyu,Maldacena:2016upp}.

\paragraph{} To see explicitly that these are zero modes of the hydrodynamic action, it is convenient to first integrate the quadratic term by parts and neglect boundary terms to obtain
\begin{equation}
\begin{aligned}
\label{eq:equiv_hydJT}
  S_{\mathrm{hydro}}^{(2)}
 =\int dt\,\biggl(&
 \left(
 2\delta\phi^{(s)}
 +a\delta g_{tt}^{(s)}
 -a\partial_t\xi_t
 \right)(\partial_t^2-r_0^2)\partial_t\xi_t
 +\frac{3ar_0^2}{8}
   {\delta g_{tt}^{(s)}}^2\\
 &-\frac{a}{4}\delta g_{tt}^{(s)}\partial_t^2\delta g_{tt}^{(s)}+\frac{r_0^2}{2}
   \delta\phi^{(s)}\delta g_{tt}^{(s)}
 -\delta\phi^{(s)}\partial_t^2
  \delta g_{tt}^{(s)}
 \biggr).
\end{aligned}
\end{equation}
This is invariant under the transformation \eqref{eq:ads2-xit-transform}, up to a boundary term $\sim \xi_t f$ that in principle can be removed by imposing suitable boundary conditions on $\xi_t$.\footnote{Note that while we have liberally neglected total derivatives in $t$ throughout, this is not strictly justified for the exponential horizon symmetry modes.} It would be worthwhile to identify analogues of the fields $(\mathcal{B},\varphi)$ for which the horizon symmetries (rather than ultraviolet diffeomorphism symmetries) are manifest and express the action in terms of these.

\paragraph{}While the constant mode $f_0$ coincides with one of the ultraviolet diffeomorphism symmetries, the exponentially growing and decaying modes do not. The three zero modes $\{f_0,f_\pm\}$ of the Schwarzian action are well-known and can be understood as inherited from the $\rm SL(2,\mathbb{R})$ isometries of AdS$_2$  \cite{Maldacena:2016upp}. What we are adding is an explicit demonstration that, at least in the linearised theory, they can also be understood as horizon symmetries along the lines anticipated in \cite{Knysh:2024asf}. To emphasise this point, in Appendix \ref{app:Dir} we repeat the calculation of the hydrodynamic action but with Dirichlet horizon boundary conditions. We find a different (non-Schwarzian) action, that reproduces the same correlators \eqref{eq:const_2D}, but has different horizon symmetries (in particular, no exponentially growing and decaying modes). 

\section{Hydrodynamic action for the BTZ black hole}
\label{sec:AdS3section}

\paragraph{}As a further example, in this Section we will derive a hydrodynamic action for the field theory dual to the BTZ black brane. As this field theory is (1+1)-dimensional, in this case the hydrodynamic action will now have two degrees of freedom: one each for temporal and spatial diffeomorphisms. As above, we use a one-sided version of the prescription of \cite{Blake:2026jue}, sufficient to obtain the retarded and advanced correlators at generic frequencies.

\paragraph{}The gravitational action is
\begin{equation}
S=\int d^3x\sqrt{-g}\left(R+2\right)+2\int_{\textrm{boundary}}d^2x\sqrt{-\gamma}\left(K+1\right).
\label{btzaction} 
\end{equation}
The boundary term is the Gibbons-Hawking term plus a holographic counterterm \cite{deHaro:2000vlm}, and is evaluated at the conformal boundary of the spacetime.\footnote{In principle there is another counterterm \cite{deHaro:2000vlm}, but this is proportional to the Ricci scalar whose integral is topological and so we will neglect it (later we will also neglect total derivatives in the boundary directions).} The equations of motion are the Einstein equations $E_{MN}=0$ where
\begin{equation}
      E_{MN}= R_{MN}-\frac{1}{2}g_{MN}\left(R+2\right).
\end{equation}
The equilibrium black hole solution can be expressed in Schwarzschild coordinates as 
\begin{equation}
\label{eq:ads3BGschwarz}
ds^2=-D(r)d\tau^2+ \frac{dr^2}{D(r)} + r^2 dx^2,
\end{equation}
where  $D(r) = r^2 - r_0^2$ and the conformal boundary is located at $r\to\infty$. The temperature and entropy density of the equilibrium black hole are related to the constant $r_0$ by $2 \pi T = r_0$ and $s = 4 \pi r_0$ respectively. We will find it convenient to work in ingoing Eddington-Finkelstein-like coordinates $(t, r, x)$ where $d\tau=dt-dr/D(r)$. In these coordinates the line element is
\begin{equation}\label{eq:BTZ_EF}
    ds^2=-D(r)dt^2 + 2 dt dr+ r^2 dx^2.
\end{equation}

\subsection*{Perturbations}

\paragraph{} We now consider linearised perturbations of the metric 
\begin{equation}
\label{eq:meticpertconventions}
    \delta ({ds^2})= \delta g_{tt} dt^2+\delta g_{xx} dx^2+\delta g_{rr} dr^2+2 \delta g_{tx} dtdx + 2 \delta g_{tr} dtdr+2 \delta g_{xr} dxdr,
\end{equation}
where $\delta g_{M N}$ are functions of $t$, $x$ and $r$. We decompose these perturbations in an analogous manner to~\eqref{metricansatz}: as a perturbation in radial gauge $h_{\mu\nu}(t,r,x)$ plus a linearised diffeomorphism $\zeta_M(t,r,x)$
    \begin{eqnarray}
    \delta g_{tt}(t,r,x) &=& - D(r) h_{tt} + 2 \partial_{t} \zeta_{t} - D'(r)(\zeta_{t} + D(r) \zeta_{r}), \nonumber \\
    \delta g_{xx}(t,r,x) &=& r^2 h_{xx} + 2 r D(r) \zeta_{r} + 2 r \zeta_{t}+ 2 \partial_{x} \zeta_{x},  \nonumber \\
    \delta g_{tx}(t,r,x) &=& r^2 h_{tx} + \partial_{t} \zeta_{x} + \partial_{x} \zeta_{t}, \nonumber \\
    \delta g_{tr}(t,r,x) &=& D'(r) \zeta_{r} + \partial_{r} \zeta_{t} + \partial_{t} \zeta_{r}, \nonumber \\    
    \delta g_{xr}(t,r,x) &=& \partial_{r} \zeta_{x} + \partial_{x} \zeta_{r} - \frac{2 \zeta_{x}}{r},  \nonumber \\
    \delta g_{rr}(t,r,x) &=& 2 \partial_{r} \zeta_{r}.
    \label{eq:htildedefn}
\end{eqnarray}
We will often find it convenient to raise indices on the diffeomorphism field via
\begin{equation}
\zeta^t=\zeta_r,\qquad\qquad
\zeta^r=\zeta_t+D(r)\zeta_r,\qquad\qquad
\zeta^x=\frac{\zeta_x}{r^2},
\label{eq:BTZ_upper_components_generic}
\end{equation}
and will assume $\zeta^M$ are regular on the future horizon. 

\subsection*{Hydrodynamic variables}

\paragraph{}Our goal is to again construct a (1+1)-dimensional hydrodynamic action
\begin{equation}
S_{\rm hydro}[\xi_t,\xi_x,\delta g^{(s)}_{\mu\nu}],
\end{equation}
where $\xi_t(t,x)$ and $\xi_x(t,x)$ are the dynamical fields (one for each conserved density $T^{t\mu}$) and $ \delta g^{(s)}_{\mu\nu}(t,x)$ are linearised external
sources for the energy-momentum tensor. As in the case of JT gravity, the
hydrodynamic fields that will naturally arise in our construction are relative temporal and spatial diffeomorphisms between the horizon and
the asymptotic boundary
\begin{eqnarray}
\xi_t(t,x) &=& [-\zeta^{t}]^{\infty}_{r_0} = -\zeta^{t}(t, r \to \infty, x) + \zeta^t(t,r_0,x), \nonumber \\
 \xi_x(t,x) &=& [\zeta^{x} - \partial_{x} \zeta^{t}/r ]_{r_0}^{\infty} = \zeta^{x}(t,r \to \infty, x) - \zeta^x(t,r_0,x) + \frac{1}{r_0} \partial_{x} \zeta^t(t,r_0,x). \label{eq:btzz_hyd}
\end{eqnarray}
These can again be written as Wilson line-like fields
\begin{equation}
    \xi_t(t,x)=-\int_{r_0}^{\infty}\frac{\delta g_{rr}}{2}dr,\qquad\qquad\xi_x(t,x)=\int^\infty_{r_0}\left(\frac{\delta g_{xr}}{r^2}-\frac{\partial_x\delta g_{rr}}{2r}\right)dr,
\end{equation}
upon integrating the final two equations in \eqref{eq:htildedefn}.

\paragraph{}To compute the hydrodynamic action we begin by expanding the gravitational action \eqref{btzaction} in powers of $\delta g_{M N}$, up to quadratic order, such that 
\begin{equation}
S = S_0 + S_1[\delta g_{M N}] + S_2[\delta g_{M N}] + \dots,
\end{equation}
with $S_1$ linear in perturbations and $S_2$ quadratic in perturbations, and the dots indicating cubic or higher terms that we ignore. We will now evaluate this action for partially on-shell solutions of the Einstein equations.

\subsection*{Equations of motion}

\paragraph{} There are six Einstein equations governing the dynamics of the small amplitude perturbations: $\delta E_{MN}=0$. These are equations for the three fields $h_{\mu\nu}$: due to gauge invariance, they do not constrain $\zeta^M$. It will be useful to separate the six equations into three dynamical equations (second order in radial derivatives) and three constraint equations (first order). We take the three dynamical equations to be
\begin{equation}
\label{eq:secondordereqsexplicit}
    \delta E_{rr} =-\frac{1}{2r^2}\partial_r\left(r^2\partial_rh_{xx}\right),\quad \delta E_{rx} = \frac{1}{2r}\partial_r\left(r^3\partial_rh_{tx}\right),\quad \delta E_{xx} = \frac{r^2}{2}\partial_r^2\left(D(r)h_{tt}\right), 
\end{equation}
and the three constraint equations to be
\begin{eqnarray}
\label{eq:firstordereqsads3}
   \text{trace} &\equiv& 2 r (\delta E_{tr} + D(r) \delta E_{rr}), \nonumber \\
     \text{momentum} &\equiv& 2 r(\delta E_{tx} + D(r) \delta E_{rx}), \nonumber \\
     \text{energy} &\equiv& - 2 r(\delta E_{tt} + D(r) \delta E_{tr}) + \frac{1}{r}\partial_x\left(\text{momentum}\right)-r \partial_t(\text{trace}).
\end{eqnarray}
The names we have given these constraints again have a simple origin: the first will ensure the field theory stress tensor obeys the Ward identity arising from Weyl transformations while the latter two ensure covariant conservation of energy and momentum in the field theory.

\paragraph{}The constraints \eqref{eq:firstordereqsads3} are radially independent once the dynamical equations \eqref{eq:secondordereqsexplicit} are imposed, due to the identities
\begin{equation}
\begin{aligned}
\label{eq:ads3EoMderivs}
\partial_r\,{\rm trace}
=&\,\frac{2}{r^2}\delta E_{xx}-\frac{2}{r}\partial_x\delta E_{rx}
-2r\left(r+\partial_t\right)\delta E_{rr},\\
\partial_r\,{\rm momentum}
=&\,-2r\partial_t\delta E_{rx}-\frac{2}{r}\partial_x\delta E_{xx},\\
\partial_r\,{\rm energy}
=&\,-\frac{2}{r^2}\left(r\partial_t+\partial_x^2\right)\delta E_{xx}-\frac{2D(r)}{r}\partial_x\delta E_{rx}+2r\left(r_0^2+r\partial_t\right)\partial_t\delta E_{rr}.
\end{aligned}
\end{equation}
Explicitly, the constraints are
\begin{equation}
\begin{aligned}
\text{trace} =&\, \partial_{r}(D(r) h_{tt}) + r^2 \partial_{r} h_{xx}-\partial_x\partial_r(r^2h_{tx})/r + \partial_{t} \partial_{r}(rh_{xx}),\\
\text{momentum} =&\, - r^3 \partial_{t} \partial_{r} h_{tx}-\partial_x\left(r D(r)\partial_{r} h_{tt}+(r^2 + r_0^2) h_{tt}\right) ,\\
\text{energy}=&\,- \partial_{t}\left((r^2 + r_0^2)  h_{tt}  +r D(r)  \partial_{r} h_{tt} +r_0^2  \partial_{r} (rh_{xx}) +r^2 \partial_{t} \partial_{r} h_{xx}\right)\\
&\, -\partial_x\left(rD(r)\partial_{r} h_{tx}-2 r_0^2h_{tx} +\partial_{x}\partial_{r}(D(r)h_{tt})\right),
\label{eq:firstordereqsexplicit}
\end{aligned}
\end{equation}
which are manifestly first order in radial derivatives.

\subsection*{Partially on-shell solution}

\paragraph{}The hydrodynamic action will be obtained from evaluating the gravitational action for partially on-shell solutions. Specifically, this means solutions for which we impose the dynamical Einstein equations \eqref{eq:secondordereqsexplicit} as well as the trace constraint in \eqref{eq:firstordereqsexplicit}.\footnote{We will later see that the off-shell stress tensor of our hydrodynamic action obeys the Weyl transformation Ward identity expected from this constraint.} 

\paragraph{}It is straightforward to integrate and find the following general solution to the dynamical equations of motion \eqref{eq:secondordereqsexplicit}
\begin{equation}
\begin{aligned}
\label{eq:firstpartialonshellsol}
D(r)h_{tt}(t,r,x)=&\, \alpha_1(t,x)r +\alpha_2(t,x), \quad\quad\quad h_{xx}(t,r,x)= \alpha_3(t,x)+\frac{\alpha_4(t,x)}{r},\\
h_{tx}(t,r,x) =&\, \alpha_5(t,x)+\frac{\alpha_6(t,x)}{r^2},
\end{aligned}
\end{equation}
where $\alpha_i(t,x)$ are six independent integration constants.\footnote{Again, the existence of simple, exact solutions to these equations is a consequence of the fact that there are no propagating gravitons in this theory and is not true for generic black holes.} For the solutions \eqref{eq:firstpartialonshellsol} the constraint equations \eqref{eq:firstordereqsexplicit} are independent of $r$ as expected from the identities \eqref{eq:ads3EoMderivs}
\begin{eqnarray}
\label{eq:conseqsalpha}
\text{trace}&=& \alpha_1-\alpha_4+\partial_t\alpha_3-2\partial_x \alpha_5, \nonumber \\
\text{momentum}&=& 2 \partial_t \alpha_6 + \partial_x\alpha_2, \nonumber \\
\text{energy} &=& \partial_{t}(\alpha_2  - r_0^2 \alpha_3 + \partial_{t} \alpha_4) + \partial_{x}(2 r_0^2 \alpha_5 + 2 \alpha_6 - \partial_{x} \alpha_1).
\end{eqnarray}
Imposing the trace constraint means fixing one relation between the six functions $\alpha_i(t,x)$. We will view this as fixing $\alpha_4$, such that our partially on-shell solution is parameterised by five independent functions $\alpha_1(t,x)$, $\alpha_2(t,x)$, $\alpha_3(t,x)$, $\alpha_5(t,x)$, $\alpha_6(t,x)$.  Importantly, we have not imposed the energy or momentum constraints yet: these will later arise as the equations of motion of the hydrodynamic action itself.

\paragraph{}For completeness, note that going partially on-shell in the way we have described is equivalent to imposing the Einstein equations $\delta E_{rr} = 0$, $\delta E_{xx} = 0$, $\delta E_{rx} = 0$ and $\delta E_{tr} = 0$, but not imposing $\delta E_{tt}=0$ or $\delta E_{tx}=0$.

\subsection*{Asymptotic boundary conditions}

\paragraph{}We are now going to fix the five remaining functions $\alpha_i(t,x)$ in the partially on-shell solutions by imposing boundary conditions on the metric at the asymptotic boundary and the horizon. As in the JT gravity case, we first expand the diffeomorphism fields near the conformal boundary as
\begin{equation}
\zeta_M(t,r,x) = \zeta_M^{(2)}(t,x) r^2+\zeta_M^{(1)}(t,x) r+\zeta_M^{(0)}(t,x) +\zeta_M^{(-1)}(t,x) r^{-1}+\ldots,
\label{diffeouv}
\end{equation}
and fix the undetermined functions in this expansion such that, upon switching to Schwarzschild coordinates \eqref{eq:ads3BGschwarz}, the partially on-shell metric perturbation is in Fefferman-Graham gauge: $\delta G_{\tau r} \sim {\mathcal O}(1/r)$, $\delta G_{rr} \sim {\mathcal O}(1/r^3)$ and $\delta G_{xr} \sim {\mathcal O}(r^{0})$. This requires\footnote{This gauge can be imposed at higher orders in the large $r$ expansion by fixing further subleading functions in \eqref{diffeouv}, although this will not be necessary for our purposes.}
\begin{equation}
\zeta_r^{(2)} = \zeta_r^{(1)} = 0, \hspace{1.5cm} \zeta_r^{(0)} = - \zeta_{t}^{(2)}, \hspace{1.5cm} \zeta_t^{(1)} = \partial_{t} \zeta_t^{(2)} .
\end{equation}
With these conditions, the remaining components of the metric perturbations in Schwarzschild coordinates are
\begin{equation}
    \begin{aligned}
        \delta G_{\tau\tau}(r\to\infty)
         &=
         -2r^2\zeta_r^{(-1)}(t,x)+O(r),
         \\
         \delta G_{\tau x}(r\to\infty)
         &=
         r^2\left(
         \alpha_5(t,x)+\partial_t\zeta_x^{(2)}(t,x)
         +\partial_x\zeta_t^{(2)}(t,x)
         \right)+O(r),
         \\
         \delta G_{xx}(r\to\infty)
         &=
         r^2\left(
         \alpha_3(t,x)+2\zeta_r^{(-1)}(t,x)
         +2\partial_t\zeta_t^{(2)}(t,x)
         +2\partial_x\zeta_x^{(2)}(t,x)
         \right)+O(r).
    \end{aligned}
\end{equation}
Using the usual holographic dictionary we identify the leading term as the perturbation of the field theory metric $\delta g_{\mu\nu}^{(s)}$ and hence 
\begin{equation}
    \begin{aligned}
    \label{eq:UV_src_BTZ}
        \delta g_{tt}^{(s)}(t,x)&\,=-2\zeta_r^{(-1)}(t,x),\\
        \delta g_{tx}^{(s)}(t,x)&\,=\alpha_5(t,x)+\partial_t\zeta_x^{(2)}(t,x)+\partial_x\zeta_t^{(2)}(t,x),\\
        \delta g_{xx}^{(s)}(t,x)&\,=\alpha_3(t,x)+2\zeta_r^{(-1)}(t,x)+2\partial_t\zeta_t^{(2)}(t,x)+2\partial_x\zeta_x^{(2)}(t,x).
    \end{aligned}
\end{equation}
We solve these equations to fix the functions $\zeta_r^{(-1)}$, $\alpha_3$ and $\alpha_5$, leaving the partially on-shell solution dependent on three undetermined functions: $\alpha_1$, $\alpha_2$ and $\alpha_6$.

\subsection*{Horizon boundary conditions}
\label{sec:IRBCs}

\paragraph{}We are now going to fix the remaining three functions by imposing boundary conditions at the horizon. The mixed horizon conditions we impose are
\begin{equation}
\delta g_{tt}(r_0)=0,\qquad
\delta g_{tx}(r_0)=0, \qquad
\delta\kappa(r_0)\equiv
\left(
\partial_r\delta g_{tt}
-2\partial_t\delta g_{tr}
+2r_0\delta g_{tr}
\right)_{r=r_0}=0, 
\label{eq:BTZ_generic_K3}
\end{equation}
where $\delta\kappa$ is the change in surface gravity.\footnote{Again these horizon boundary conditions, and the remainder of the calculation of the hydrodynamic action, simplify when working in radial gauge near the horizon as we show in Appendices \ref{app:radial-gauge-actions} and \ref{app:radial-gauge-horizon-symmetries}.} As illustrated in Appendix~\ref{gen_bcs_JT-new}, these are consistent with the variational principle for the quadratic gravitational action at the horizon (in the absence of a Gibbons-Hawking term). This ensures that the equations of motion of the hydrodynamic action we obtain will be the remaining Einstein constraint equations that we have not yet imposed.  

\paragraph{} The boundary conditions \eqref{eq:BTZ_generic_K3} fix the remaining three functions to
\begin{eqnarray}
\alpha_1(t,x) &=& -2\zeta^r(t,r_0,x) - 2 r_0\partial_t \zeta^t(t,r_0,x) - 2 \partial_t^2 \zeta^t(t,r_0,x), \nonumber \\
\alpha_2(t,x) &=& 2\partial_t\zeta^r(t,r_0,x) +2r_0^2\partial_t \zeta^t(t,r_0,x) +2r_0\partial_t^2\zeta^t(t,r_0,x),
\nonumber \\
\alpha_6(t,x) &=&-r_0^2\alpha_5(t,x) -r_0^2\partial_t\zeta^x(t,r_0,x) -\partial_x\zeta^r(t,r_0,x),
\label{eq:BTZ_generic_alpha}
\end{eqnarray}
where $\zeta^M(t,r_0,x)$ are the horizon diffeomorphisms introduced earlier.

\paragraph{}After imposing the relations \eqref{eq:BTZ_generic_alpha}, the horizon diffeomorphisms combine with asymptotic diffeomorphisms such that the unimposed Einstein constraints become equations for the relative diffeomorphisms $\xi_t(t,x)$ and $\xi_x(t,x)$ defined in \eqref{eq:btzz_hyd}. After defining
\begin{equation}
B_{tx}
=
\delta g^{(s)}_{tx}
-\partial_t\xi_x
-\partial_x\xi_t, \quad B_{xx}
=
\delta g^{(s)}_{xx}
-2\partial_x\xi_x, \quad B_{tt}
=
\delta g^{(s)}_{tt}
-2\partial_t\xi_t, \label{eq:BTZ_Blin_generic}
\end{equation}
the energy and momentum constraints can be written explicitly as
\begin{equation}
{\rm momentum}
=
-2r_0^2\partial_t B_{tx}, \qquad {\rm energy}
=
\partial_t\left((\partial_t^2-r_0^2)(B_{tt}+B_{xx})
-2\partial_t\partial_xB_{tx}\right).
\label{eq:BTZ_mom_ene_generic}
\end{equation}
We will soon see that these are the equations of motion of the hydrodynamic action.

\subsection*{Linear hydrodynamic action}

\paragraph{} We will now obtain the hydrodynamic action by evaluating the gravitational action for the partially on-shell solutions. At linear order in the perturbation amplitude we find 
\begin{equation}
S_{\rm hydro}^{(1)}[\xi_\mu, \delta g_{\mu\nu}^{(s)}]
=
\int d^2x
\left(
\frac{r_0^2}{2}\delta g^{(s)}_{tt}
+
\frac{r_0^2}{2}\delta g^{(s)}_{xx}
\right).
\label{eq:Slinearterm}
\end{equation}
where we have integrated by parts and neglected terms that are total derivatives in $t$ and $x$ (as we will do throughout).
From the linear action we can read off the equilibrium energy density $\langle T^{tt}_{\textrm{eq}} \rangle = r_0^2$ and pressure $\langle T^{xx}_{\textrm{eq}} \rangle = r_0^2$. 

\subsection*{Quadratic hydrodynamic action}

\paragraph{} At quadratic order in the perturbation amplitude, the action is more complicated. To proceed, we first integrate the (off-shell) Einstein-Hilbert term by parts so
that
\begin{equation}
S_{\rm EH}=S_{\rm bulk}+S_{\rm boundary}, \quad {\rm with} \quad S_{\rm bulk}
=
\int d^3x\,\frac r2\,\delta E_{MN}\delta g^{MN}.
\label{eq:BTZ_EHsplit_generic}
\end{equation}
We will return to $S_{\rm boundary}$ shortly. With our parameterisation of the metric perturbations, the bulk term is
\begin{align}
S_{\rm bulk}
&=
\frac12
\int d^3x
\left(
-\frac1r\delta E_{xx}h_{xx}
+rD(r)\delta E_{rr}h_{tt}
-2r\delta E_{tx}h_{tx}
\right. \nonumber\\
&\left.+\partial_r
\left(
\zeta_r\times{\rm energy}
+
\left(
-\frac{\zeta_x}{r^2}
+\frac{\partial_x\zeta_r}{r}
\right)
\times{\rm momentum}
-
\left(
D(r)\zeta_r+\zeta_t+r\partial_t\zeta_r
\right)
\times{\rm trace}
\right)
\right).
\label{eq:BTZ_bulk_before_onshell_generic}
\end{align}

\paragraph{} Our partially on-shell solutions satisfy the three Einstein equations on the first line of \eqref{eq:BTZ_bulk_before_onshell_generic}, as well as the trace constraint, and so for these solutions the bulk action collapses to the boundary
term
\begin{equation}
S_{\rm bulk}
=
\frac12
\int d^3x\,\partial_r
\left(
\zeta_r\times{\rm energy}
+
\left(
-\frac{\zeta_x}{r^2}
+\frac{\partial_x\zeta_r}{r}
\right)
\times{\rm momentum}
\right).
\label{eq:BTZ_bulk_boundary_generic}
\end{equation}
Since the energy and momentum constraints are radially conserved for our partially on-shell solutions, this boundary term can be expressed simply in terms of the relative diffeormorphisms between the boundary and horizon
\begin{equation}
S_{\rm bulk}
=
-\frac12
\int d^2x
\left(
\xi_t\times{\rm energy}
+
\xi_x\times{\rm momentum}
\right),
\label{eq:BTZ_Sbulk_generic}
\end{equation}
where the relative diffeormorphisms $\xi_\mu$ are the hydrodynamic degrees of freedom defined in \eqref{eq:btzz_hyd}.  Finally, substituting in the explicit form of the energy and momentum constraints~\eqref{eq:BTZ_mom_ene_generic} gives
\begin{equation}
S_{\rm bulk}
=
-\frac12
\int d^2x
\left(
\xi_t\partial_t
\left(
(\partial_t^2-r_0^2)(B_{tt}+B_{xx})
-2\partial_t\partial_xB_{tx}
\right)
-2r_0^2\xi_x\partial_tB_{tx}
\right).
\label{eq:BTZ_Sbulk_B_generic}
\end{equation}

\paragraph{} We now return to $S_{\textrm{boundary}}$, which in principle has contributions from both the asymptotic boundary and the horizon. In Appendix \ref{app:BTZhorizonbdyterm} we evaluate the horizon contribution and show that it vanishes for the partially on-shell solution constructed with the horizon boundary conditions above. At the asymptotic boundary, there are contributions from $S_{\textrm{boundary}}$ as well as the Gibbons-Hawking term and counterterm in equation \eqref{btzaction}. Evaluating these together for our partially on-shell solution gives
\begin{equation}
S^{(2)}_{\textrm{UV bdy}}
=
\frac12
\int d^2x\left((\delta g^{(s)}_{tt}+\delta g^{(s)}_{xx})
(\partial_t^2-r_0^2)\partial_t\xi_t
-
2
\delta g^{(s)}_{tx}\partial_t
\left(
r_0^2\xi_x
+
\partial_t\partial_x\xi_t
\right)\right)
+
S^{(2)}_{\rm source}[\delta g^{(s)}_{\mu\nu}],
\label{eq:BTZ_SUV_generic}
\end{equation}
where $S^{(2)}_\textrm{source}[\delta g_{\mu\nu}^{(s)}]$ are the following quadratic source terms that are independent of the hydrodynamic fields 
\begin{equation}
\begin{aligned}
S^{(2)}_{\rm source}[\delta g^{(s)}_{\mu\nu}]
=
\int d^2x
\Bigg(&
\frac{r_0^2}{2}
{\delta g^{(s)}_{tx}}^2
-\frac{r_0^2}{8}
{\delta g^{(s)}_{xx}}^2
+\frac{3r_0^2}{8}
{\delta g^{(s)}_{tt}}^2+\frac{r_0^2}{4}
\delta g^{(s)}_{tt}\delta g^{(s)}_{xx}\\
&
-\frac{1}{4}
\delta g^{(s)}_{tt}(\partial_t^2+\partial_x^2)\delta g^{(s)}_{tt}
-\partial_x\delta g^{(s)}_{tx}\partial_t\delta g^{(s)}_{tt}
-\frac{1}{2}
\delta g^{(s)}_{tt}\partial_t^2\delta g^{(s)}_{xx}
\Bigg). \label{eq:BTZ_contact_generic}
\end{aligned}
\end{equation}

\paragraph{}Finally, combining with $S_{\textrm{bulk}}$ in \eqref{eq:BTZ_Sbulk_B_generic}, the complete quadratic hydrodynamic action is
\begin{align}
S^{(2)}_{\rm hydro}
&=
\int d^2x
\Big(
\xi_t(\partial_t^2-r_0^2)\partial_t^2\xi_t
-\xi_t\partial_t^2\partial_x^2\xi_t
-r_0^2\xi_x\partial_t^2\xi_x
-2r_0^2\xi_t\partial_t\partial_x\xi_x \nonumber\\
&+
(\delta g^{(s)}_{tt}+\delta g^{(s)}_{xx})
(\partial_t^2-r_0^2)\partial_t\xi_t
-
2\delta g^{(s)}_{tx}
\partial_t\left(
r_0^2\xi_x
+
\partial_t\partial_x\xi_t
\right)
\Big)
+
S^{(2)}_{\rm source}[\delta g^{(s)}_{\mu\nu}].
\label{eq:BTZ_hydro_action_generic}
\end{align}
In the absence of sources, the hydrodynamic action simplifies to
\begin{equation}
\label{eq:BTZ-action-nosources}
S_{\rm hydro}[\xi_\mu]=
\int d^2x
\left(
\xi_t(\partial_t^2-\partial_x^2-r_0^2)\partial_t^2\xi_t-r_0^2\xi_x\partial_t^2\xi_x
-2r_0^2\xi_t\partial_t\partial_x\xi_x
\right).   
\end{equation}

\paragraph{}In chiral coordinates $x^\pm=x\pm t$, the action in the absence of sources is
\begin{equation}\label{eq:non-fac_BTZ}
    S_{\rm hydro}=2\int dx^+dx^- \begin{pmatrix} \xi_+ & \xi_- \end{pmatrix}\begin{pmatrix}-(\partial_-\partial_t+r_0^2)\partial_+\partial_t & \partial_+\partial_-\partial_t^2 \\ \partial_+\partial_-\partial_t^2 & -(\partial_+\partial_t-r_0^2)\partial_-\partial_t\end{pmatrix}\begin{pmatrix} \xi_+ \\ \xi_- \end{pmatrix}.
\end{equation}
Unlike the geometric Virasoro actions of \cite{Haehl:2018izb,Cotler:2018zff,Haehl:2019eae,Nguyen:2021jja,Nguyen:2021pdz,Nguyen:2025jhc}, this does not factorize holomorphically. However it is diffeomorphism invariant, like the action on the plane obtained in Section 3 of \cite{Nguyen:2021pdz}. This is due to us renormalising the holographic theory with the standard, diffeomorphism-invariant counterterms given in \eqref{btzaction}.\footnote{See Section 2 of \cite{Nguyen:2021pdz} for a discussion of the counterterm needed to enforce factorization at the expense of diffeomorphism invariance.}

\subsection*{Compact form of the quadratic action}

\paragraph{}Although the hydrodynamic action looks rather complicated, it can be simplified (and some of its symmetries made manifest) by writing it in terms of perturbations of the induced metric under a diffeomorphism $\mathcal{B}_{\mu\nu}(t,x)$ that was defined in equation \eqref{eq:curly_B}. To quadratic order in amplitudes, this is 
\begin{equation}
\mathcal B_{\mu\nu}
  =
  \delta g^{(s)}_{\mu\nu}
  -\partial_\mu\xi_\nu
  -\partial_\nu\xi_\mu
  -\xi^\rho\partial_\rho\delta g^{(s)}_{\mu\nu}
  -\delta g^{(s)}_{\rho\nu}\partial_\mu\xi^\rho
  -\delta g^{(s)}_{\mu\rho}\partial_\nu\xi^\rho
  +\partial_\mu\xi^\alpha\partial_\nu\xi_\alpha,
  \label{bmunu}
\end{equation}
where indices on $\xi_\mu$ are raised and lowered with the Minkowski metric. Note that to linear order in perturbation amplitudes $\mathcal{B}_{\mu\nu}$ coincide with the objects $B_{\mu \nu}$ previously introduced in \eqref{eq:BTZ_Blin_generic}. 

\paragraph{}In terms of these fields the hydrodynamic action can be expressed more compactly as
\begin{equation}\label{eq:BTZ_hydro_action}
\begin{aligned}
S_{\mathrm{hydro}}[\xi_\mu,\delta g_{\mu\nu}^{(s)}]= \frac{1}{2}\int d^{2} x \Biggl(&r_0^2\left({\mathcal B}_{tt}+{\mathcal B}_{xx}+\mathcal{B}_{tx}^2-\frac{1}{4}\mathcal{B}_{xx}^2+\frac{3}{4}\mathcal{B}_{tt}^2+\frac{1}{2}\mathcal{B}_{tt}\mathcal{B}_{xx}\right)\\
&\,-\frac{1}{2}\mathcal{B}_{tt}(\partial_t^2+\partial_x^2)\mathcal{B}_{tt}-2\partial_x\mathcal{B}_{tx}\partial_t\mathcal{B}_{tt}-\mathcal{B}_{tt}\partial_t^2\mathcal{B}_{xx}\Biggr),
\end{aligned}
\end{equation}
up to quadratic order in perturbation amplitudes. Note that this may be written as
\begin{equation}
S_{\mathrm{hydro}}[\xi_\mu,\delta g^{(s)}_{\mu \nu}] =  \frac{r_0^2}{2}\int d^{2} x \left({\mathcal B}_{tt}+{\mathcal B}_{xx}\right) +  S^{(2)}_{\mathrm{source}}[{\mathcal B}_{\mu \nu}].  \label{BTZ_pullback_action}
\end{equation}

\paragraph{}To make the Weyl transformation properties of the action manifest, we define
\begin{equation}
    \mathcal{G}_{\mu\nu}(x^\mu)=\eta_{\mu\nu}+\mathcal{B}_{\mu\nu}(x^\mu)= \frac{\partial y^{\alpha}}{\partial x^{\mu} } \frac{\partial y^{\beta}}{\partial x^{\nu}}  g^{(s)}_{\alpha \beta}(y(x^\mu)),
\end{equation}
where $y^\mu=x^\mu-\xi^\mu$, and then the fields
\begin{equation}
    \Psi(x^\mu)\equiv\log(-\mathcal{G}_{tt}(x^\mu)),\quad\quad\quad\quad\quad\quad
    \tilde{\mathcal{G}}_{\mu\nu}(x^\mu)\equiv e^{-\Psi(x^\mu)}\mathcal{G}_{\mu\nu}(x^\mu).
\end{equation}
While both fields are diffeomorphism invariant, only the field $\tilde{\mathcal{G}}_{\mu\nu}(x^\mu)$ is invariant under the Weyl transformation $g_{\mu\nu}^{(s)}(y(x^\mu))\rightarrow e^{-2\Omega(y(x^\mu))}g_{\mu\nu}^{(s)}(y(x^\mu))$ while $\Psi(x^\mu)\rightarrow\Psi(x^\mu)-2\Omega(y(x^\mu))$ shifts. To quadratic order in amplitudes, they are explicitly given by
\begin{equation}
\begin{aligned}
\tilde{\mathcal{G}}_{tt}&\,=-1,\quad\quad\quad\quad\quad\quad\quad\quad\quad\quad \,\tilde{\mathcal{G}}_{xx}=1+\mathcal{B}_{tt}+\mathcal{B}_{xx}+\mathcal{B}_{tt}(\mathcal{B}_{tt}+\mathcal{B}_{xx}),\\
\tilde{\mathcal{G}}_{tx}&\,=\mathcal{B}_{tx}(1+\mathcal{B}_{tt}),\quad\quad\quad\quad\quad\quad\quad\Psi=-\mathcal{B}_{tt}-\frac{1}{2}\mathcal{B}_{tt}^2,\\
\end{aligned}    
\end{equation}
with $\mathcal{B}_{\mu\nu}$ defined in equation \eqref{bmunu}.

\paragraph{}To quadratic order, the hydrodynamic action can then be written\footnote{The equilibrium contribution of this action $r_0^2\int dtdx$ gives the correct value of the thermal partition function of a 2D CFT ($4\pi^2 T$ in our units) upon analytic continuation to Euclidean signature.}
\begin{equation}
\label{eq:ads3hydroactioncompact}
    S_{\text{hydro}}=\frac{1}{2}\int d^2x\sqrt{-\tilde{\mathcal{G}}}\left(2r_0^2+\frac{1}{2}\tilde{\mathcal{G}}^{\mu\nu}\partial_\mu\Psi\partial_\nu\Psi+\Psi R[\tilde{\mathcal{G}}]\right), 
\end{equation}
where $R[\tilde{\mathcal{G}}]$ denotes the Ricci scalar of $\tilde{\mathcal{G}}$. The Weyl anomaly of the action is isolated in the final two terms, which has the form of the Liouville action although with $\Psi$ not the fundamental field (in this sense, our action is related to that of \cite{Nguyen:2021pdz}). The first term can be understood as a consequence of evaluating the Liouville action around the non-trivial scalar solution appropriate for the thermal state rather than the vacuum. Similarly to the JT gravity case, the first term in the Lagrangian can also be understood as a covariant expression for the thermodynamic pressure
\begin{equation}
\int dt\sqrt{-g^{(s)}}P[g^{(s)}].\quad\quad\quad\quad P[g^{(s)}]=4\pi^2\left(-g_{\mu\nu}^{(s)}\beta^\mu\beta^\nu\right)^{-1}, 
\end{equation}
with $g^{(s)}$ replaced by $\mathcal{G}$. 

\subsubsection*{Constitutive relations, equations of motion and two-point functions} 

\paragraph{}We are now going to use the explicit hydrodynamic action to obtain expressions for the energy-momentum tensor and show that putting the hydrodynamic fields on-shell ensures that this tensor is covariantly conserved. We will further show that the two-point functions obtained from the action are those of a thermal 2d CFT.

\paragraph{} We define the energy-momentum tensor $T^{\mu\nu}$ as
\begin{equation}\label{eq:BTZ_st_defn}
T^{\mu \nu} \equiv \frac{2}{\sqrt{-g^{(s)}}} \frac{\delta S_{\textrm{hydro}}}{\delta (\delta g_{\mu \nu}^{(s)})}.
\end{equation}
The equilibrium values were given after equation \eqref{eq:Slinearterm}, and evaluating it explicitly to linear order in perturbations gives\footnote{Before differentiating, one must replace $\delta g_{tx}^{(s)}$ by $\frac{1}{2}(\delta g_{xt}^{(s)}+\delta g_{tx}^{(s)})$ everywhere in the action since our notation implicitly assumes the metric is symmetric.}
\begin{equation}
\begin{split}
\label{eq:taudefn}
  \delta   T^{tt} &= 2\left(\partial_t^2 - r_0^2 \right)\partial_t\xi_t+\left(2 r_0^2 -\partial_t^2-\partial_x^2\right)\delta g_{tt}^{(s)}-\partial_t^2\delta g_{xx}^{(s)}+2\partial_x\partial_t\delta g_{tx}^{(s)},  \\
    \delta T^{xx} &= 2\left(\partial_t^2 - r_0^2 \right)\partial_t\xi_t+\left(r_0^2 -\partial_t^2\right)\delta g_{tt}^{(s)}- r_0^2 \delta g_{xx}^{(s)},  \\
     \delta T^{tx} &= -2 r_0^2 \partial_t\xi_x - 2\partial_x\partial_t^2\xi_t+ r_0^2 \delta g_{tx}^{(s)}+\partial_x\partial_t\delta g_{tt}^{(s)},
\end{split}
\end{equation}
where we have used $\sqrt{-g}=1+\frac{1}{2}(\delta g_{xx}^{(s)}-\delta g_{tt}^{(s)})$. These are off-shell expressions, and can be thought of as constitutive relations expressing stress tensor perturbations in terms of the hydrodynamic fields and external sources. 

\paragraph{}The off-shell energy-momentum tensor identically satisfies
\begin{equation}
\label{eq:ads3weyl}
    -\delta T^{tt}+\delta T^{xx}+r_0^2(\delta g_{tt}^{(s)}+\delta g_{xx}^{(s)})=\partial_x^2\delta g_{tt}^{(s)}+\partial_t^2\delta g_{xx}^{(s)}-2\partial_x\partial_t\delta g_{tx}^{(s)}.
\end{equation}
This is the linearization, around the equilibrium state, of the Ward identity for Weyl transformations in a 2D CFT $T^\mu_{\;\;\mu}=R[g^{(s)}]$ where the conformal anomaly on the right hand side is the Ricci scalar of the metric $g^{(s)}$.

\paragraph{}To progress further we must put the hydrodynamic fields on shell. Varying the action yields the equations of motion
\begin{equation}
\begin{split}
   \label{eq:eomshydro}
   0 &=\frac{\delta S_\text{hydro}}{\delta \xi_t}= -\partial_t\left((\partial_t^2-r_0^2)(B_{tt}+B_{xx})
-2\partial_t\partial_xB_{tx}\right),\\   
     0&=\frac{\delta S_\text{hydro}}{\delta \xi_x}= 2 r_0^2 \partial_{t} B_{tx},
\end{split}
\end{equation}
which (up to an overall minus sign) are precisely the Einstein constraint equations~\eqref{eq:BTZ_mom_ene_generic} that we did not yet impose. In other words, after integrating out the hydrodynamic fields we are studying solutions to the full set of Einstein equations. In terms of the energy-momentum tensor \eqref{eq:taudefn} these equations of motion are
\begin{equation}
    \begin{aligned}
    \label{eq:ads3diffeo}
           0 &=\frac{\delta S_\text{hydro}}{\delta \xi_t}= \partial_t\left(\delta T^{tt}+r_0^2\delta g_{xx}^{(s)}-r_0^2\delta g_{tt}^{(s)}\right)+\partial_x\left(\delta T^{tx}-r_0^2\delta g_{tx}^{(s)}\right),\\   
     0&=\frac{\delta S_\text{hydro}}{\delta \xi_x}= \partial_{t}\left(\delta T^{tx}+r_0^2\delta g_{tx}^{(s)}\right)+\partial_{x}\left(\delta T^{xx}+r_0^2\delta g_{xx}^{(s)}-r_0^2\delta g_{tt}^{(s)}\right).
    \end{aligned}
\end{equation}
These are just the linearisation of the covariant conservation equations $\nabla_\mu T^{\mu\nu}=0$ around the equilibrium state, and so integrating out the hydrodynamic fields enforces these conservation laws.

\paragraph{}After a Fourier transformation, the equations of motion can be solved explicitly for the hydrodynamic fields, except for modes with frequencies $\omega=0$, $\omega^2=-r_0^2$ and $\omega^2=k^2$. The solutions are:
\begin{equation}\label{eq:freq_3D}
\begin{aligned}
\xi_t(\omega,k)
&=
\frac{i}{2(\omega^2-k^2)}
\left[
\omega\left(
\delta g^{(s)}_{tt}(\omega,k)
+\delta g^{(s)}_{xx}(\omega,k)
\right)
+2k\,\delta g^{(s)}_{tx}(\omega,k)
\right],
\\
\xi_x(\omega,k)
&=
\frac{i}{2(\omega^2-k^2)}
\left[
k\left(
\delta g^{(s)}_{tt}(\omega,k)
+\delta g^{(s)}_{xx}(\omega,k)
\right)
+2\omega\,\delta g^{(s)}_{tx}(\omega,k)
\right].
\end{aligned}
\end{equation}
After putting these on-shell, we then define the two-point functions via functional derivatives of the energy momentum tensor
\begin{equation}
\label{eq:ads32ptfndefn}
G_{T^{\mu \nu} T^{\rho \sigma}}(\omega, k)  = 2 \frac{\delta (\sqrt{-g^{(s)}} T^{\mu \nu})}{\delta g_{\rho \sigma}^{(s)}}.
\end{equation}
Explicitly, this gives\footnote{The two-point functions defined in this way are symmetric upon interchanging the two operators.}
\begin{equation}
\begin{aligned}
\label{eq:ads32ptfns}
G_{T^{tt} T^{tt}}(\omega, k) =&  \frac{2k^2(k^2+r_0^2)}{k^2 - \omega^2}+r_0^2,\quad\quad\quad\qquad\;\;\;\;\;
G_{T^{tx} T^{tx}}(\omega, k) =  \frac{2 \omega^2 (k^2+r_0^2)}{k^2 - \omega^2}+r_0^2,\\
G_{T^{xx} T^{xx}}(\omega, k) =&  \frac{2k^2(\omega^2+r_0^2)}{k^2 - \omega^2}-2\omega^2-3r_0^2,\quad\quad\quad G_{T^{tt} T^{tx}}(\omega, k) =   \frac{2 \omega k(k^2 + r_0^2)}{k^2 - \omega^2},\\
G_{T^{tt} T^{xx}}(\omega, k) =&   \frac{2 \omega^2 (k^2+r_0^2)}{k^2 - \omega^2}+r_0^2,\quad\qquad\quad\quad\;\;\;\;\;
G_{T^{tx} T^{xx}}(\omega, k) = \frac{2 \omega k(\omega^2 + r_0^2)}{k^2 - \omega^2},
\end{aligned}
\end{equation}
which can equivalently be obtained from the on-shell hydrodynamic action via
\begin{equation}
S_{\text{hydro, on-shell}} =\frac{1}{8} \int \frac{d\omega dk}{(2 \pi)^2}  \delta g_{\mu \nu}^{(s)}(-\omega, -k) G_{T^{\mu \nu} T^{\rho \sigma}}(\omega, k) \delta g^{(s)}_{\rho \sigma}(\omega, k).
\end{equation}
These should again be understood as the retarded (or advanced) Green's functions at generic frequencies, and are equivalent to those obtained by replacing $\omega_E\rightarrow\pm i\omega$ in the Euclidean correlators. In the thermal state of a 2D CFT, the two-point functions of the stress tensor are fixed entirely by the Weyl and diffeomorphism Ward identities \eqref{eq:ads3weyl} and \eqref{eq:ads3diffeo}. Since our energy-momentum tensor satisfies these on-shell, \eqref{eq:ads32ptfns} manifestly agree with the correct CFT expressions.\footnote{We note again that other definitions of the Green's functions are sometimes used in the literature, which differ from these by contact terms.} An $i\epsilon$ prescription is necessary to deal with poles at real $\omega$. As for JT gravity, implementing this is necessary to obtain the symmetric correlator and it would be worthwhile to understand how to do this directly in the gravitational derivation of the action.

\subsection*{Ultraviolet symmetries}

\paragraph{}Having explicitly confirmed that our hydrodynamic action produces the right correlation functions, we will now return and examine its symmetries. There are again ultraviolet symmetries related to asymptotic diffeomorphisms, and horizon symmetries related to horizon diffeomorphisms.

\paragraph{}As for JT gravity, the ultraviolet symmetries of the action ensure that the hydrodynamic equations of motion are the conservation equations \eqref{eq:ads3diffeo}, and that the off-shell stress tensor is traceless up to the conformal anomaly \eqref{eq:ads3weyl}. To see the former, consider $t$ and $x$ diffeomorphisms that preserve Fefferman-Graham gauge near the asymptotic boundary and vanish near the horizon. These act to shift the hydrodynamic fields $\xi^\mu(x^\mu)\rightarrow\xi^\mu(x^\mu)+\epsilon^\mu(x^\mu-\xi^\mu(x^\mu))$ with a compensating transformation of $g_{\mu\nu}^{(s)}$ such that $\mathcal{B}_{\mu\nu}$ is invariant. Explicitly, to linear order in $\epsilon(x^\mu)$ and linear order in the perturbation amplitude, these transformations are
\begin{equation}
\begin{aligned}
    &\,\delta_\epsilon\xi_t=-\epsilon^t-\xi_t\partial_t\epsilon^t+\xi_x\partial_x\epsilon^t,\quad\quad\quad\quad \delta_\epsilon\xi_x=\epsilon^x+\xi_t\partial_t\epsilon^x-\xi_x\partial_x\epsilon^x, \\
&\,\delta_\epsilon\delta g_{tt}^{(s)}=-2\partial_t\epsilon^t+\epsilon^x\partial_x\delta g_{tt}^{(s)}+\epsilon^t\partial_t\delta g_{tt}^{(s)}+2\delta g_{tt}^{(s)}\partial_t\epsilon^t+2\delta g_{tx}^{(s)}\partial_t\epsilon^x,\\
&\,\delta_\epsilon\delta g_{xx}^{(s)}=2\partial_x\epsilon^x+\epsilon^x\partial_x\delta g^{(s)}_{xx}+\epsilon^t\partial_t\delta g_{xx}^{(s)}+2\delta g_{xx}^{(s)}\partial_x\epsilon^x+2\delta g_{tx}^{(s)}\partial_x\epsilon^t,\\
&\,\delta_\epsilon\delta g_{tx}^{(s)}=-\partial_x\epsilon^t+\partial_t\epsilon^x+\delta g_{tt}^{(s)}\partial_x\epsilon^t+\delta g_{xx}^{(s)}\partial_t\epsilon^x+\partial_t(\delta g_{tx}^{(s)}\epsilon^t)+\partial_x(\delta g_{tx}^{(s)}\epsilon^x).
\end{aligned}
\end{equation}
The action written in the form \eqref{eq:BTZ_hydro_action} or \eqref{eq:ads3hydroactioncompact} manifestly satisfies $\delta_\epsilon S_{\text{hydro}}=0$, which ensures its equations of motion are the Ward identities \eqref{eq:ads3diffeo}.

\paragraph{}Asymptotic radial diffeomorphisms $\chi^r=-r\Omega(t,x)+\ldots$ (with subleading terms in $\chi^M$ to preserve Fefferman-Graham gauge) act as Weyl transformations on the sources
\begin{equation}
\begin{aligned}
    \delta_\Omega\delta g_{tt}^{(s)}&\,=2(1-\delta g_{tt}^{(s)})\Omega,\quad\quad\quad\quad\quad\quad\delta_\Omega\delta g_{xx}^{(s)}=-2(1+\delta g_{xx}^{(s)})\Omega,\\
    \delta_\Omega\delta g_{tx}^{(s)}&\,=-2\delta g_{tx}^{(s)}\Omega,
    \end{aligned}
\end{equation}
to linear order in perturbation amplitude. Keeping $\xi_\mu$ fixed under these yields $\delta_\Omega\tilde{\mathcal{G}}_{\mu\nu}=0$ and $\delta_\Omega\Psi=-2\Omega-2\xi_t\partial_t\Omega+2\xi_x\partial_x\Omega$, with which it is straightforward (using the form of the action given in equation \eqref{eq:ads3hydroactioncompact}) to compute that the action transforms anomalously as  $\delta S/\delta\Omega=-\sqrt{-g^{(s)}}R[g^{(s)}]$ to the order we are working, where $R[g^{(s)}]$ denotes the Ricci scalar of the metric $g^{(s)}$. Rewriting the left hand side in terms of the off-shell hydrodynamic stress tensor, we see that the Weyl transformation properties of the action ensure that the trace Ward identity \eqref{eq:ads3weyl} is obeyed.

\subsection*{Horizon symmetries}

\paragraph{}We now examine the horizon symmetries which arise from diffeomorphisms that vanish near the boundary and preserve the horizon boundary conditions \eqref{eq:BTZ_generic_K3} that we imposed when deriving the hydrodynamic action. We parameterise the horizon diffeomorphism as $\chi(t,r_0,x)=f(t,x)\partial_t+Z(t,x)\partial_r+\tilde{X}(t,x)\partial_x$, which acts on the horizon metric perturbations as $\delta g_{MN}\rightarrow\delta g_{MN}+\nabla_M\chi_N+\nabla_N\chi_M$. In our hydrodynamic action, it leaves the sources $\delta g_{\mu\nu}^{(s)}$ invariant (to linear order in perturbations) while shifting the hydrodynamic fields as
\begin{equation}    \delta_H\xi_t=f(t,x),\quad\quad\quad\quad\quad\delta_H\xi_x=-\tilde{X}(t,x)+\frac{1}{r_0}\partial_xf(t,x),
\end{equation}
at zeroeth order.\footnote{Since these fields only enter the action at quadratic order (up to total derivatives), this form of the transformation is sufficient.} It acts on the quantities appearing in our horizon boundary conditions \eqref{eq:BTZ_generic_K3} as
\begin{equation}
\begin{aligned}
    \delta_H(\delta g_{tt}(t,r_0,x)) &= 2\partial_t Z-2r_0Z, \quad\quad\quad\quad\quad\quad  \delta_H(\delta g_{tx}(t,r_0,x)) &= r_0^2\partial_t\widetilde X+\partial_x Z,\\
\delta_H(\delta\kappa(t,r_0,x))  &=   -2 Z -2 r_0\partial_t f - 2 \partial_t^2 f,
\label{eq:BTZ_hor_sym_fXZ}
\end{aligned}
\end{equation}
to zeroeth order in perturbation amplitude.

\paragraph{}The horizon diffeomorphisms that leave the boundary conditions \eqref{eq:BTZ_generic_K3} unchanged can easily be found by integrating the right hand side of \eqref{eq:BTZ_hor_sym_fXZ} to get
\begin{equation}
\begin{aligned}
f(t,x)
&\,=
a_2(x)+a_4(x)e^{-r_0t}
-\frac{a_3(x)}{2r_0^2}e^{r_0t},\quad\quad\quad\quad\quad Z(t,x)=a_3(x)e^{r_0t},\\
\tilde{X}(t,x)&\,=a_5(x)
-\frac{\partial_x a_3(x)}{r_0^3}e^{r_0t}.
\end{aligned}
\end{equation}
After a redefinition of the arbitrary functions of $x$, the hydrodynamic fields transform under these as 
\begin{equation}
    \begin{aligned}
    \label{eq:ads3horizonhydrotransforms}
        &\,\delta_H\xi_t=c_0(x)+c_-(x)e^{-r_0t}
+c_+(x)e^{r_0t},\\
&\,\delta_H\xi_x=c_x(x)
+\frac{\partial_xc_-(x)}{r_0}e^{-r_0t}-\frac{\partial_x c_+(x)}{r_0}e^{r_0t},
    \end{aligned}
\end{equation}
The time dependence of these are the same as in JT gravity: there are constant and exponentially growing and decaying modes. It is straightforward to check that both $\mathcal{B}_{\mu\nu}$ and $\tilde{\mathcal{G}}_{\mu\nu}$ transform non-trivially under \eqref{eq:ads3horizonhydrotransforms} and so these horizon symmetries are different from the asymptotic diffeomorphisms or Weyl transformations discussed above.

\paragraph{}We again interpret the modes
\begin{equation}
\begin{aligned}
\label{eq:ads3hydrozeromodes}
    \xi_t(t,x)&\,=c_0(x)+c_-(x)e^{-r_0t}+c_+(x)e^{r_0t},\\
    \xi_x(t,x)&\,=c_x(x)+\frac{\partial_xc_-(x)}{r_0}e^{-r_0t}-\frac{\partial_xc_+(x)}{r_0}e^{r_0t},
    \end{aligned}
\end{equation}
as redundancies of our parameterisation. It is straightforward to check the off-shell constitutive relations for the stress tensor \eqref{eq:taudefn} are invariant under adding such modes, as are the hydrodynamic equations of motion \eqref{eq:eomshydro}. This means that these modes carry no energy-momentum, although they are expected to be important for chaotic properties of the state.\footnote{Note that there are additional non-trivial solutions to the hydrodynamic equations in the absence of sources that do carry energy-momentum.}

\paragraph{}Finally, to see that \eqref{eq:ads3hydrozeromodes} are true zero modes of the hydrodynamic action, it is convenient to first integrate \eqref{eq:BTZ_hydro_action_generic} by parts and write it in the form\footnote{With the same caveat as in JT gravity that we are not being careful with the total derivative terms, even for exponential growing and decaying modes.}
\begin{align}
S_{\mathrm{hydro}}^{(2)}
=\int d^2x\,\Biggl(&
\left(
\delta g_{tt}^{(s)}
+\delta g_{xx}^{(s)}
-\partial_t\xi_t
-\partial_x\xi_x
\right)
(\partial_t^2-r_0^2)\partial_t\xi_t
\nonumber\\
&+
\left(
\partial_x\xi_t
+\partial_t\xi_x
-2\delta g_{tx}^{(s)}
\right)
\left(
r_0^2\partial_t\xi_x
+\partial_x\partial_t^2\xi_t
\right)
\nonumber\\
&+\frac{r_0^2}{2}{\delta g_{tx}^{(s)}}^2
-\frac{r_0^2}{8}{\delta g_{xx}^{(s)}}^2
+\frac{3r_0^2}{8}{\delta g_{tt}^{(s)}}^2
+\frac{r_0^2}{4}\delta g_{tt}^{(s)}\delta g_{xx}^{(s)}
\nonumber\\
&-\frac14\delta g_{tt}^{(s)}
(\partial_t^2+\partial_x^2)\delta g_{tt}^{(s)}
-\partial_x\delta g_{tx}^{(s)}
 \partial_t\delta g_{tt}^{(s)}
-\frac12\delta g_{tt}^{(s)}
 \partial_t^2\delta g_{xx}^{(s)}
\Biggr).
\label{eq:BTZ_IBP}
\end{align}
This is invariant under the transformations \eqref{eq:ads3horizonhydrotransforms} even to quadratic order in the zero modes, up to a boundary term that can in principle be removed by an appropriate choice of boundary conditions on $\xi_\mu$. This fact is not obvious from the forms of the action we have presented, and it would be worthwhile to identify horizon analogues of the variables $\mathcal{B}_{\mu\nu}$ which make the horizon symmetries manifest.

\subsection*{$\xi_\mu$ two-point functions}

\paragraph{}A key motivation for us in identifying horizon symmetries of hydrodynamic action is their purported importance for chaotic out-of-time-ordered correlations \cite{Blake:2017ris}. In order to compare with the formalism of \cite{Blake:2017ris}, we now briefly analyse the two-point functions of the hydrodynamic fields $\xi_\mu$ themselves (see \cite{Haehl:2018izb} for a similar analysis of a different hydrodynamic action for a 2D CFT).

\paragraph{}After a Fourier transformation, the hydrodynamic action in the absence of sources \eqref{eq:BTZ-action-nosources} is
\begin{align}
S_{\rm hydro}
=
\int {d\omega\,dk\over (2\pi)^2}
\begin{pmatrix}
\xi_t(-\omega,-k) & \xi_x(-\omega,-k)
\end{pmatrix}
\begin{pmatrix}
\omega^2(\omega^2+r_0^2-k^2) & -r_0^2\omega k\\
-r_0^2\omega k & r_0^2\omega^2
\end{pmatrix}
\begin{pmatrix}
\xi_t(\omega,k)\\
\xi_x(\omega,k)
\end{pmatrix}.
\label{eq:BTZ_xi_matrix}
\end{align}
Inverting the matrix gives the following momentum space propagators for the hydrodynamic fields
\begin{equation}
\begin{split}
&\langle \xi_t(\omega,k)\xi_t(-\omega,-k)\rangle
 =
 {1\over(\omega^2+r_0^2)(\omega^2-k^2)}
,\\
&\langle \xi_t(\omega,k)\xi_x(-\omega,-k)\rangle
 =
 \langle \xi_x(\omega,k)\xi_t(-\omega,-k)\rangle
 =
 {k\over
 \omega(\omega^2+r_0^2)(\omega^2-k^2)}, \\
&\langle \xi_x(\omega,k)\xi_x(-\omega,-k)\rangle
 =
 {\omega^2+r_0^2-k^2\over
 r_0^2(\omega^2+r_0^2)(\omega^2-k^2)}.
 \label{eq:BTZ_xit_xit}
\end{split} 
\end{equation}
The form of these two-point functions is very similar to the structure obtained in~\cite{Blake:2017ris} for maximally chaotic theories with a single hydrodynamical field. In particular, a Fourier transform of (any) of these two-point functions, as described in \cite{Blake:2017ris,Haehl:2018izb}, includes exponentially growing modes of the form $\sim e^{r_0(t - |x|)}$ (due to the pole at $\omega = i r_0$). This is precisely the functional form of the out-of-time-ordered correlators of few-body operators in this theory, as anticipated in \cite{Blake:2017ris}.

\paragraph{}The matrix in \eqref{eq:BTZ_xi_matrix} is non-invertible for the frequencies $\omega=0,\pm ir_0,\pm k$. For the frequencies $\omega=\pm ir_0$ corresponding to the exponential horizon symmetries, the null eigenvectors are $\xi_x=\mp ik\xi_t/r_0$. These are the momentum space form of the exponential horizon symmetry modes we identified in \eqref{eq:ads3hydrozeromodes} above. 

\section{Discussion} \label{sec:disc}

\paragraph{}We have derived exact hydrodynamic actions for black holes in JT and AdS$_3$ gravity, for small amplitude perturbations around the equilibrium state. We followed the prescription of \cite{Blake:2026jue}, in which the hydrodynamic degrees of freedom arise as relative diffeomorphisms between the asymptotic boundary and the horizon. This prescription requires a choice of horizon boundary conditions, which affects the action one obtains. Motivated by the variational principle in the absence of a horizon Gibbons-Hawking term, we mainly focused on the boundary conditions in equation \eqref{ads23bcsi}. With this choice, the resulting actions are given in equations \eqref{quadomegaaction} and \eqref{eq:ads3hydroactioncompact} for JT and AdS$_3$ respectively. The former reduces to the Schwarzian action of \cite{Maldacena:2016upp} when external sources vanish.

\paragraph{}We have shown that the hydrodynamic actions exhibit symmetries naturally associated with the black hole horizon. Diffeomorphisms that preserve the horizon boundary conditions (and vanish near the asymptotic boundary) correspond to transformations of the hydrodynamic fields that leave the action invariant. For the boundary conditions \eqref{ads23bcsi}, the horizon symmetries are exactly those proposed in \cite{Knysh:2024asf} and conjectured there to be related to symmetries of hydrodynamic actions. Our explicit construction makes this connection precise.

\paragraph{}There is not a unique hydrodynamic action that yields the correct stress tensor two-point functions. In particular we showed that imposing Dirichlet boundary conditions on the (stretched) horizon, after including a Gibbons-Hawking term there, yields a different action that produces the same stress tensor two-point functions. This action also naturally has different horizon symmetries. This is not inconsistent. The horizon symmetries are redundancies of how the hydrodynamic action is parameterised: the physical stress tensor is invariant under them.

\paragraph{}However, the boundary conditions used in the main text are noteworthy as they support a proposed connection between hydrodynamics and OTOCs. Specifically, it was argued in \cite{Blake:2017ris} that the $\sim e^{2 \pi T(t - |x|/v_B)}$ profile of OTOCs can be understood as arising from operators coupling to a hydrodynamic action with pointwise exponentially growing zero modes. The actions in the main text contain such modes and we have argued that similar pointwise shifts can arise in  higher dimensional cases \cite{Blake:2026jue}. The coupling of operators to similar hydrodynamic degrees of freedom has been studied in \cite{Maldacena:2016hyu,Haehl:2018izb} for JT and AdS$_3$ gravity, and it would be very interesting to extend this to higher dimensions (see \cite{Ammon:2025vod} for work in this direction). For the forms of the hydrodynamic action without such a zero mode (obtained with Dirichlet boundary conditions), it would be interesting to understand whether the OTOC can be computed from an effective theory with a different coupling between operators and the hydrodynamic degrees of freedom than that of \cite{Blake:2017ris}.

\paragraph{} We have focused in this paper on linear perturbations about thermal states, since in the higher dimensions where \cite{Blake:2026jue} was developed this is a significant practical simplification. However the examples discussed in this paper are particularly simple, and a full non-linear analysis should be possible. An important question is to identify the full non-linear generalisation of the boundary conditions~\eqref{ads23bcsi} that give rise to hydrodynamic actions which realise the horizon symmetries of~\cite{Knysh:2024asf}. There are a number of technical improvements which could also be made to our calculation. Firstly, by working covariantly throughout it should be possible to directly arrive on the compact diffeomorphism and Weyl-covariant actions (equations~\eqref{quadomegaaction} and \eqref{eq:ads3hydroactioncompact}), rather than reassembling these from more explicit, but complicated, expressions. Secondly, it would be worthwhile understanding how the $i\epsilon$  regularisation can be implemented directly in the gravitational derivation of the action, rather than at the level of correlation functions.\footnote{As explained in Section \ref{sec:AdS2section} this likely requires carefully considering dynamics on the full CGL contour for the $\omega=0$ mode, rather than just one branch of it.} Thirdly, it is important to determine exactly what (temporal) boundary conditions on hydrodynamic fields are allowed, and use these to systematically deal with the total $\partial_t$ derivatives that arise in the Lagrangian upon integration by parts. Finally, we imposed a particular Einstein constraint equation to obtain the hydrodynamic action. It would be interesting to see how the action changes if a linear combination of the constraints was instead imposed.

\paragraph{}Our discussion of horizon symmetries in Section~\ref{sec:hor_sy} was presented for a one-sided black hole with a single set of hydrodynamic degrees of freedom. This is sufficient for the actions we then obtained. For more general black holes, there are two sets of hydrodynamic degrees of freedom (reflecting the two branches of the Schwinger-Keldysh contour) that are coupled through horizon boundary conditions. Nevertheless, one can still identify horizon symmetries as residual diffeomorphisms preserving the horizon boundary conditions. For the case of the AdS$_4$-Schwarzschild black hole, studied in detail in \cite{Blake:2026jue}, the horizon symmetries identified in this way again contain pointwise exponential modes (for a natural choice of boundary conditions). Determining whether these are realised as symmetries of an action for the stress tensor dynamics requires determining this action non-perturbatively in derivatives. Unlike the cases studied here, in general this non-perturbative action will be non-local beyond the thermal scale.

\acknowledgments
\paragraph{} We are grateful to Akash Jain, Hong Liu, M\'{a}rk Mezei and Natalia Pinzani-Fokeeva for helpful discussions. MB acknowledges support from UK Research and Innovation (UKRI)
under the UK government’s Horizon Europe guarantee (EP/Y00468X/1). AD acknowledges support from a recently concluded STFC Consolidated Grant ST/T000600/1 -- ``Particle Theory at the Higgs Centre''. The work of AD is supported by the grant ANRF/ECRG/2024/000247/PMS from the Anusandhan National Research Foundation (ANRF), India. The work of RD was partially supported by the STFC Ernest Rutherford Grant ST/R004455/1. There is no underlying data associated to this research. For the purpose of open access, the authors have applied a Creative Commons Attribution (CC BY) licence to any Author Accepted Manuscript version arising from this submission.

\appendix

\section{Quadratic action variational principle}\label{gen_bcs_JT-new}
\paragraph{}In this Appendix we motivate the horizon boundary conditions we used in deriving the hydrodynamic actions in the main text. We explain why with these boundary conditions the equations of motion of the hydrodynamic action are the constraint Einstein equations that were not imposed when deriving the action.

\subsection*{JT gravity}

\paragraph{}We first consider JT gravity, in particular the quadratic in amplitudes piece of the bulk (off-shell) Einstein-Hilbert action. Under a variation of the metric and scalar field perturbation
\begin{equation}\label{eq:var_met2}
\delta g_{M N} \to \delta g_{M N}+ \delta \chi_{M N}, \hspace{1.0cm} \delta \Phi \to  \delta \Phi+ \delta \Psi,
\end{equation}
the variation in the bulk action can be written as 
\begin{equation}
\delta S_{\mathrm{EH}}= \int d^2 x\bigg[\delta E_{\Phi} \delta \Psi + \delta E_{M N}
 \delta \chi^{M N}\bigg] +  S_{\mathrm{bdy}}(\infty) - S_{\mathrm{bdy}}(r_0), 
\end{equation}
to linear order in $(\delta\chi_{MN},\delta\Psi)$ where $\delta E_{\Phi}$ and $\delta E_{M N}$ are the linearised Einstein equations for $\delta g_{MN}$ and $\delta \Phi$, and the boundary term is 
\begin{equation}
\begin{aligned}
 S_{\mathrm{bdy}} =&\int dt \Biggl(- \frac{3}{2} a r^2 D(r)^2 \delta g_{rr} \delta \chi_{rr} - 4 a r^2 \delta g_{tr} \delta \chi_{tr} + a \delta g_{tr} \delta \chi_{tt} - a r \delta g_{tt,r} \delta \chi_{tr}   \\
 &- a r \delta g_{tr} \delta \chi_{tt,r}-   \delta \Phi_{,r} \delta \chi_{tt} + \delta \Phi  \delta \chi_{tt,r} + r D(r)  \delta \Phi \delta \chi_{rr}- 3 a r^2 D(r) \delta g_{rr} \delta \chi_{tr}   \\
 &-  3 a r^2 D(r) \delta g_{tr} \delta \chi_{rr}- a rD(r) \delta g_{rr,t} \delta \chi_{tr}- a rD(r) \delta g_{tr}\delta \chi_{rr,t}- 2 \delta \Phi \delta \chi_{tr,t}   \\
&  + 2 r \delta \Phi  \delta \chi_{tr}-\frac{a}{2} (r^2 + r_0^2)  \delta g_{rr} \delta \chi_{tt}- a r^2  \delta g_{tt} \delta \chi_{rr}-  \frac{1}{2} a r  \delta g_{tt} \delta \chi_{rr,t}       \\
&+  \frac{1}{2} a r  \delta g_{rr} \delta \chi_{tt,t}- \frac{a r}{2} D(r) \delta g _{rr} \delta \chi_{tt,r}   - \frac{a r}{2} D(r) \delta g_{tt,r}    \delta \chi _{rr}\Biggr), 
\end{aligned}
\end{equation}
where commas denote partial derivatives with respect to the subsequent coordinate.

\paragraph{} At the asymptotic boundary, this boundary term is supplemented by the variation of the Gibbons-Hawking term, and the total boundary term vanishes under Dirichlet boundary conditions (i.e. fixing the boundary metric and scalar source). 

\paragraph{}In the main text we do not add a Gibbons-Hawking term at the horizon. Recalling that all fields are regular on the horizon
\begin{eqnarray}
S_{\mathrm{bdy}}(r_0) &=& \int dt \bigg(-4  a r_0^2 \delta g_{tr} \delta \chi_{tr} + a \delta g_{tr} \delta \chi_{tt} - a r_0 \delta g_{tt, r} \delta \chi_{tr}   - a r_0 \delta g _{tr} \delta \chi_{tt,r}  - a r_0^2  \delta g_{tt} \delta \chi_{rr} \nonumber \\ 
 &-&  a r_0^2  \delta g_{rr} \delta \chi_{tt}  - \frac{a r_0}{2} \delta g_{tt} \delta \chi_{rr,t} + \frac{a r_0}{2} \delta g_{rr} \delta \chi_{tt,t}  - \partial_{r} \delta \Phi \delta \chi_{tt}  +2 r_0 \delta \Phi  \delta \chi_{tr} \nonumber \\
 &+&  \delta \Phi \partial_{r} \delta \chi_{tt} - 2 \delta \Phi \partial_{t} \delta \chi_{tr} \bigg)_{r=r_0}. \nonumber
\end{eqnarray}
For this to vanish (up to total derivatives in $t$) it is sufficient to impose the boundary conditions
\begin{equation}
\begin{split}
&\delta g_{tt} = 0, \hspace{1.0cm}   \delta\kappa \equiv \partial_{r} \delta g_{tt} - 2 \partial_{t} \delta g_{tr} + 2 r_0 \delta g_{tr} = 0, \\
&\delta \chi_{tt} = 0, \hspace{1.0cm} \delta\kappa_\chi \equiv \partial_{r} \delta \chi_{tt} - 2 \partial_{t} \delta \chi_{tr} + 2 r_0 \delta \chi_{tr} = 0, 
\end{split}\label{gen_h_bcs}
\end{equation}
at the horizon $r=r_0$.\footnote{Note that if working in radial gauge near the horizon fixing the values of $\delta g_{tt}$ and $\partial_r\delta g_{tt}$ on the horizon is sufficient to ensure $S_{\mathrm{bdy}}(r_0)=0$.} Fixing $\delta g_{tt}$ and $\delta\kappa$ in this way are the horizon boundary conditions \eqref{eq:generic-gauge-horizon-bcs} that we impose in the main text.

\paragraph{} The vanishing of $S_{\mathrm{bdy}}(r_0)$ is what ensures that the equation of motion of our hydrodynamical action is equivalent to the Einstein equation that was left unimposed when going partially on-shell. In particular, after imposing the dynamical Einstein equations and the trace constraint the bulk contribution to the linear variation can be expressed as
\begin{eqnarray}
\int d^2 x\left(\delta E_{\Phi} \delta \Psi + \delta E_{M N}
 \delta \chi^{M N}\right) &=&   \int d^{2} x \partial_{r} \left(\delta \zeta^{t} \times \mathrm{energy} \right)  \nonumber \\
 &=& - \int dt \left( \delta \xi_t \times \mathrm{energy} \right),
 \label{eq:JTvariapp}
\end{eqnarray}
where we have parameterised $\delta \chi_{MN}$ in an analogous manner to~\eqref{metricansatz} as 
\begin{equation}
\begin{split}
\delta \chi_{tt}(t, r)&= -D(r) \delta h_{tt}(t, r)+ 2\partial_t\delta \zeta_t(t,r) - D'(r)\big(\delta \zeta_t(t,r)+ D(r)\delta \zeta_r(t,r)\big), \\
\delta \chi_{tr}(t,r) &=  D'(r)\delta \zeta_r(t,r) + \partial_r\delta \zeta_t(t,r) +  \partial_t\delta \zeta_r(t,r), \\
\delta \chi_{rr}(t, r) &= 2\partial_r \delta \zeta_r(t,r),
\end{split}
\end{equation}
and identified $\delta \xi_t = -\delta \zeta^{t}(\infty) + \delta \zeta^{t}(r_0)$ as the variation of the hydrodynamic mode. Therefore provided we impose Dirichlet boundary conditions at the asymptotic boundary, and $S_{\mathrm{bdy}}(r_0)=0$, the total variation of the partially on-shell action is \eqref{eq:JTvariapp}. Since the partially on-shell action is the hydrodynamic action, the hydrodynamic equation of motion is the energy constraint.

\subsection*{BTZ black hole}

\paragraph{}We now repeat this analysis of the variational principle for the BTZ black hole. We consider a variation of the quadratic (in $\delta g_{MN})$ part of the Einstein-Hilbert action under $\delta g_{MN} \to \delta g_{MN} + \delta \chi_{MN}$. To linear order in $\chi_{MN}$ we again have
\begin{equation}
 \delta S_{EH}
 =
 \int d^3x\,r\delta E_{MN}\delta \chi^{MN}
 +S_{\rm bdy}(\infty)-S_{\rm bdy}(r_0),
 \label{eq:BTZ_full_variation_split}
\end{equation}
where $\delta E_{MN}$ are the linearised Einstein equations evaluated on $\delta g_{MN}$ and $S_{\rm bdy}$ is a boundary term. At the asymptotic boundary there is an additional variation of the Gibbons-Hawking term and counterterm, and the combination vanishes after fixing the boundary metric in the usual way.  

\paragraph{}In the main text we do not add a Gibbons-Hawking term at the horizon, and want to identify boundary conditions such that the variation $S_{\rm bdy}(r_0)$ vanishes. Explicitly, this term is
\begin{equation}
\begin{aligned}
S_{\rm bdy}(r_0)
&=\int d^2x
\Bigg[
\left(
    \frac{\delta g_{xx}}{4r_0}-r_0 \delta g_{tr}
\right)\delta\kappa_\chi
+
\left(
    \frac{\delta \chi_{xx}}{4r_0}-r_0\delta \chi_{tr}
\right)\delta\kappa+\delta g_{tt}
\biggl(
 -\frac{3}{2}\delta \chi_{tr}
\\
& -r_0^2\delta \chi_{rr}
 -\frac{r_0}{4}\partial_t\delta \chi_{rr}
 -\frac{1}{r_0}\partial_x\delta \chi_{rx}
 +\frac{3}{4r_0}\partial_r\delta \chi_{xx}
 -\frac{3}{4r_0^2}\delta \chi_{xx}
\biggr)
\\
&
+\delta \chi_{tt}
\left(
 -\frac{3}{2}\delta g_{tr}
 -r_0^2 \delta g_{rr}
 -\frac{r_0}{4}\partial_t \delta g_{rr}
 -\frac{1}{r_0}\partial_x \delta g_{rx}
 +\frac{3}{4r_0}\partial_r \delta g_{xx}
 -\frac{3}{4r_0^2}\delta g_{xx}
\right)
\\
&
+\frac{r_0}{4}
\left(
    (\partial_t \delta g_{tt})\delta \chi_{rr}
    +(\partial_t\delta \chi_{tt})\delta g_{rr}
\right)
-\frac{1}{2r_0}
\left(
    (\partial_x \delta g_{tt})\delta \chi_{rx}
    +(\partial_x\delta \chi_{tt})\delta g_{rx}
\right)
\\
&
+\delta g_{tx}
\left(
 \frac{1}{r_0}\partial_x\delta \chi_{tr}
 -\frac{3}{2r_0}\partial_r\delta \chi_{tx}
 -2\delta \chi_{rx}
 +\frac{1}{2r_0}\partial_t\delta \chi_{rx}
 -\frac{1}{4r_0^3}\partial_x\delta \chi_{xx}
\right)
\\
&
+\delta \chi_{tx}
\left(
 \frac{1}{r_0}\partial_x \delta g_{tr}
 -\frac{3}{2r_0}\partial_r \delta g_{tx}
 -2\delta g_{rx}
 +\frac{1}{2r_0}\partial_t \delta g_{rx}
 -\frac{1}{4r_0^3}\partial_x \delta g_{xx}
\right)
\\
&
+\frac{1}{r_0}
\left(
    (\partial_t \delta g_{tx})\delta \chi_{rx}
    +(\partial_t\delta \chi_{tx})\delta g_{rx}
\right)-(\partial_x \delta g_{tx})
\left(
    \frac{1}{2r_0}\delta \chi_{tr}
    +\frac{1}{4r_0^3}\delta \chi_{xx}
\right)
\\
&-(\partial_x\delta \chi_{tx})
\left(
    \frac{1}{2r_0}\delta g_{tr}
    +\frac{1}{4r_0^3}\delta g_{xx}
\right)
\Bigg]_{r=r_0},
\label{eq:BTZ_horizon_variation_bilinear}
\end{aligned}
\end{equation}
with $\delta\kappa$ and $\delta\kappa_\chi$ defined as in~\eqref{gen_h_bcs}.

\paragraph{} The horizon boundary conditions we imposed in the main text are 
\begin{equation}
 \delta g_{tt}=0,\quad
 \delta g_{tx}=0,\quad
 \delta\kappa=0,
 \qquad (r=r_0),
 \label{eq:BTZ_background_horizon_bcs_appA}
\end{equation}
and the allowed variations preserving them satisfy
\begin{equation}
 \delta \chi_{tt}=0,\quad
 \delta \chi_{tx}=0,\quad
 \delta\kappa_\chi=0,
 \qquad (r=r_0).
 \label{eq:BTZ_variation_horizon_bcs_appA}
\end{equation}
Since $\delta g_{tt}$, $\delta g_{tx}$, $\delta \chi_{tt}$ and $\delta \chi_{tx}$ vanish as functions on the horizon, their tangential $(t, x)$ derivatives also vanish there, as then does $S_{\mathrm{bdy}}(r_0)$. Hence the quadratic Einstein-Hilbert action has a well-defined variational principle at the horizon (without adding a Gibbons-Hawking term there) if the boundary conditions \eqref{eq:BTZ_background_horizon_bcs_appA} and \eqref{eq:BTZ_variation_horizon_bcs_appA} are imposed.

\paragraph{} As for JT gravity, the consistency of this variational principle ensures that the equations of motion of the hydrodynamic action are equivalent to the unimposed Einstein equations (the energy and momentum constraints). The bulk contribution to the linear variation, after imposing the dynamical equations of motion and trace constraint, becomes
\begin{eqnarray}
\int d^{3}x \left(r\delta E_{MN} \delta {\chi}^{MN}\right) &=& \int d^{2} x \partial_{r}\left( \delta \zeta_{r} \times \mathrm{energy} + \bigg(-\frac{\delta \zeta_x}{r^2} + \frac{\partial_{x} \zeta_{r}}{r} \bigg) \times \mathrm{momentum} \right) \nonumber \\
&=& - \int d^{2}x \left(\delta \xi_{t} \times \mathrm{energy} + \delta \xi_{x} \times \mathrm{momentum}  \right),
\end{eqnarray}
where we have parameterised $\delta \chi_{MN}$ as
    \begin{eqnarray}
    \delta \chi_{tt}(t,r,x) &=& - D(r) \delta h_{tt} + 2 \partial_{t} \delta \zeta_{t} - D'(r)(\delta \zeta_{t} + D(r) \delta \zeta_{r}), \nonumber \\
    \delta \chi_{tr}(t,r,x) &=& D'(r) \delta \zeta_{r} + \partial_{r} \delta \zeta_{t} + \partial_{t} \delta \zeta_{r}, \nonumber \\
    \delta \chi_{rr}(t,r,x) &=& 2 \partial_{r} \delta \zeta_{r}, \nonumber \\
    \delta \chi_{tx}(t,r,x) &=& r^2 \delta h_{tx} + \partial_{t} \delta \zeta_{x} + \partial_{x} \delta \zeta_{t}, \nonumber \\
    \delta \chi_{xr}(t,r,x) &=& \partial_{r} \delta \zeta_{x} + \partial_{x} \delta \zeta_{r} - \frac{2 \delta \zeta_{x}}{r}, \nonumber \\
    \delta \chi_{xx}(t,r,x) &=& r^2 \delta h_{xx} + 2 r D(r) \delta \zeta_{r} + 2 r \delta \zeta_{t}+ 2 \partial_{x}\delta \zeta_{x}, 
\end{eqnarray}
and identified the variation of the hydrodynamic modes as $\delta \xi_{t} = [-\delta \zeta^{t}]^{\infty}_{r_0}$ and $\delta \xi_x = [\delta \zeta^{x} - \partial_{x} \delta \zeta^{t}/r]^{\infty}_{r_0}$. We thus see that with the horizon boundary conditions above, and asymptotic Dirichlet boundary conditions, variations of the partially on-shell action (i.e.~the hydrodynamic action) with respect to $\xi_t$ and $\xi_x$ give the unimposed Einstein constraint equations (up to an overall sign). 

\section{Boundary term for quadratic action in BTZ spacetime}
\label{app:BTZhorizonbdyterm}

\paragraph{}In this Appendix we show that the horizon contribution to $S_{\textrm{boundary}}$ for the BTZ black hole vanishes with the horizon boundary conditions \eqref{eq:BTZ_generic_K3}. This fact was used after equation \eqref{eq:BTZ_Sbulk_B_generic} in the main text derivation of the hydrodynamic action.

\paragraph{}Explicitly, the horizon contribution can be written
\begin{equation}
\begin{aligned}
S_{\rm boundary}&(r_0)
=
\int d^2x
\Bigg(
\left(
    \frac{\delta g_{xx}}{4r_0}-r_0 \delta g_{tr}
\right)\delta\kappa+\frac{r_0}{4}(\partial_t \delta g_{tt})\delta g_{rr}
-\frac{1}{2r_0}(\partial_x \delta g_{tt})\delta g_{rx}\\
&
+\delta g_{tt}
\left(
 -\frac{3}{2}\delta g_{tr}
 -r_0^2 \delta g_{rr}
 -\frac{r_0}{4}\partial_t \delta g_{rr}
 -\frac{1}{r_0}\partial_x \delta g_{rx}
 +\frac{3}{4r_0}\partial_r \delta g_{xx}
 -\frac{3}{4r_0^2}\delta g_{xx}
\right)\\
&
+\delta g_{tx}
\left(
 \frac{1}{r_0}\partial_x \delta g_{tr}
 -\frac{3}{2r_0}\partial_r \delta g_{tx}
 -2\delta g_{rx}
 +\frac{1}{2r_0}\partial_t \delta g_{rx}
 -\frac{1}{4r_0^3}\partial_x \delta g_{xx}
\right)\\
&
+\frac{1}{r_0}(\partial_t \delta g_{tx})\delta g_{rx}
-(\partial_x \delta g_{tx})
\left(
    \frac{1}{2r_0}\delta g_{tr}
    +\frac{1}{4r_0^3}\delta g_{xx}
\right)
\Bigg)_{r=r_0}.
\label{eq:BTZ_quadratic_horizon_boundaryterm_appA}
\end{aligned}
\end{equation}
Imposing the horizon boundary conditions~\eqref{eq:BTZ_generic_K3} and using the fact that the $t$ and $x$ derivatives of $\delta g_{tt}$ and $\delta g_{tx}$ also vanish
on the horizon, this reduces to a total $t$ derivative (which we neglect throughout)
\begin{equation}
S_{\rm boundary}(r_0)
=
\int d^2x\,\partial_t
\left(
 -\frac{r_0}{2}(\delta g_{tr})^2
 +\frac{1}{4r_0}\delta g_{tr}\delta g_{xx}
 +\frac{1}{2r_0^3}(\delta g_{xx})^2
\right)_{r=r_0}.
\label{eq:BTZ_action_horizon_total_derivative_appA}
\end{equation}

\section{Near-horizon radial gauge calculations of hydrodynamic action}
\label{app:radial-gauge-actions}

\paragraph{}In this Appendix we briefly outline how the calculations of the hydrodynamic actions in the main text are modified if one enforces radial gauge near the horizon (rather than allowing $\zeta^M$ to be a generic regular perturbation there). This approach yields the same answer but is technically simpler.

\subsection*{JT gravity}

\paragraph{} We continue working with perturbations of the background \eqref{eq:EF_JT} and now impose radial gauge (in ingoing coordinates) near $r=r_0$ i.e. we want $\delta g_{tr}$ and $\delta g_{rr}$ to vanish there. Specifically, this means restricting the form of the diffeomorphism fields near the horizon to
\begin{eqnarray}\label{eq:resid_2D}
\zeta_r(t, r \to r_0, x)&=&T(t) + \mathcal{O}((r - r_0)^2), \nonumber \\
\zeta_t(t, r \to r_0, x) &=&-D(r)T(t)-r\partial_tT(t) + R(t)  + \mathcal{O}((r - r_0)^2),
\end{eqnarray}
where $T(t)$ and $R(t)$ parameterise the residual diffeomorphisms that preserve radial gauge near the horizon. 

The analysis in the main text from~\eqref{expandaction} to~\eqref{UVbc} goes through completely unchanged. However restricting to radial gauge near the horizon simplifies the boundary conditions and expression for the integration constants $\alpha_1$ and $\alpha_2$. Specifically, in radial gauge the horizon boundary conditions become (using \eqref{eq:generic-gauge-horizon-bcs})
\begin{equation}\label{eq:rad_gag_cndJT}
\delta g_{tt}(r_0)=0,\qquad\qquad\qquad\qquad\qquad
\partial_r\delta g_{tt}(r_0)=0 .
\end{equation}
Restricting the gauge fields to the form~\eqref{eq:resid_2D} these boundary conditions fix the integration constants $\alpha_1$ and $\alpha_2$ to
\begin{equation}\label{eq:al12_rad_JT}
\alpha_1=-2R-2\partial_t^2T,\qquad\qquad\qquad\qquad\qquad
\alpha_2=2r_0^2\partial_tT+2\partial_tR.
\end{equation}
Inserting these expressions into the energy constraint~\eqref{eq:bulk_energy} gives 
\begin{equation}
\begin{split}
\mathrm{energy}=
(\partial_t^2 - r_0^2)\partial_t
\left(
2\delta\phi^{(s)}
+a\left(\delta g^{(s)}_{tt}-2\partial_t\xi_t\right)
\right)= 0, 
\end{split}
\end{equation}
which is the same as the corresponding equation~\eqref{energy_explicit} in the main text, but now where
\begin{equation}
\label{eq:xi_t_radial_JT}
\xi_t(t) = [-\zeta^{t}]^{\infty}_{r_0} = \zeta^{(2)}_t(t) + T(t).
\end{equation}

\paragraph{} The rest of the calculation proceeds in the same manner in the main text. This expression for the energy constraint can be used to compute the bulk contribution to the action. The asymptotic boundary terms are unchanged, and the boundary term~\eqref{boundary1} vanishes at the horizon after imposing radial gauge and the boundary conditions~\eqref{eq:rad_gag_cndJT}. The final result for the hydrodynamic action is again~\eqref{hydroJT}, where the hydrodynamic field is given by the relative diffeomorphism~\eqref{eq:xi_t_radial_JT}.

\subsection*{BTZ black hole}

\paragraph{}For the BTZ black hole case, imposing radial gauge  $\delta g_{tr} = \delta g_{rr} = \delta g_{xr}=0$ near the horizon requires that the diffeomorphisms are
\begin{eqnarray}
\label{eq:app_BTZ_res_radg}
\zeta_r(t,r \to r_0, x) &=& T(t,x) + \mathcal{O}((r-r_0)^2), \nonumber \\
\zeta_x(t, r \to r_0, x)  &=&r^2 X(t,x)+r\partial_xT(t,x) + \mathcal{O}((r-r_0)^2), \nonumber   \\
\zeta_t(t, r \to r_0, x) &=&-D(r)T(t,x)-r\partial_tT(t,x) + R(t, x) + \mathcal{O}((r-r_0)^2). 
\end{eqnarray}
There is now a three-parameter family of residual diffeomorphisms that preserve radial gauge, specified by the independent functions $T(t,x)$, $R(t,x)$ and $X(t,x)$. The explicit expressions for the hydrodynamic fields become
\begin{equation}
\begin{aligned}
    \label{wil_BTZ}
    \xi_t(t,x)&\,=[-\zeta^{t}]^{\infty}_{r_0} = \zeta_t^{(2)}(t,x)+T(t,x),\\
    \xi_x(t,x)&\,= [\zeta^{x} -\partial_{x} \zeta^{t}/r]^{\infty}_{r_0} = \zeta_x^{(2)}(t,x)-X(t,x).
    \end{aligned}
\end{equation}
The analysis through to~\eqref{eq:UV_src_BTZ} is unchanged, but working in radial gauge near the horizon modifies the expressions for the integration constants $\alpha_1$, $\alpha_2$ and $\alpha_6$. The horizon boundary conditions \eqref{eq:BTZ_generic_K3} reduce to
\begin{equation}
\delta g_{tt}(r_0)=0,\qquad\qquad
\partial_r\delta g_{tt}(r_0)=0,\qquad\qquad
\delta g_{tx}(r_0)=0 .
\end{equation}
The integration constants are then fixed to
\begin{equation}\label{eq:al126_BTZ}
\alpha_1 =-2R-2\partial_t^2T, \quad
\alpha_2 =2r_0^2\partial_tT+2\partial_tR, \quad
\alpha_6 =-r_0^2\alpha_5-\partial_xR-r_0^2\partial_tX .
\end{equation}
With these boundary conditions, the energy and momentum constraints have the same form as~\eqref{eq:BTZ_mom_ene_generic} with $\xi_\mu$ given by the explicit expressions~\eqref{wil_BTZ}. The rest of the calculations proceed through unchanged, and one again obtains~\eqref{eq:BTZ_hydro_action_generic} for the hydrodynamic action with $\xi_\mu$ given by \eqref{wil_BTZ}.

\section{Horizon symmetries as residual diffeomorphisms} \label{app:radial-gauge-horizon-symmetries}

\paragraph{}Following on from the previous Appendix, where we calculated the hydrodynamic actions by working in radial gauge near the horizon, in this Appendix we show that the horizon symmetries of these actions can be interpreted as residual gauge transformations that preserve radial gauge near the horizon in addition to the horizon boundary conditions.

\subsection*{JT gravity}
\paragraph{}For JT gravity, consider acting on metric perturbations $\delta g_{MN}$ with a diffeomorphism that preserves radial gauge near the horizon. This can be achieved by considering a diffeomorphism $\delta g_{MN} \to \delta g_{MN} + \nabla_{N} \chi_{M} + \nabla_{M} \chi_{N}$ where
\begin{eqnarray}\label{eq:resid_3D}
\chi_r(t, r \to r_0, x)&=&\delta T(t) + \mathcal{O}((r - r_0)^2), \nonumber \\
\chi_t(t, r \to r_0, x) &=&-D(r)\delta T(t)-r\partial_t \delta T(t) + \delta R(t)  + \mathcal{O}((r - r_0)^2).
\end{eqnarray}
The fields $\delta T$ and $\delta R$ can be viewed as shifts of the residual diffeomorphism fields in~\eqref{eq:rad_gag_cndJT}. Under such a diffeomorphism one has
\begin{eqnarray}
\delta g_{tt}(r_0) &\to& \delta g_{tt}(r_0) - 2 r_0 \delta R + 2 r_0^2 \partial_{t} \delta T + 2 \partial_{t} \delta R - 2 r_0 \partial_{t}^2 \delta T, \nonumber \\
\partial_{r} \delta g_{tt}(r_0) &\to& \partial_{r} \delta g_{tt}(r_0) - 2 \delta R - 2 \partial_{t}^2 \delta T.
\end{eqnarray}
The horizon boundary conditions $\delta g_{tt}(r_0) = \partial_{r} \delta g_{tt}(r_0) = 0$ are therefore preserved by residual diffeomorphisms of the form 
\begin{equation}\label{eq:rad_JT_hor}
\delta T(t)=c_0+c_+e^{r_0t}+c_-e^{-r_0t},
\qquad
\delta R(t)=-r_0^2c_+e^{r_0t}-r_0^2c_-e^{-r_0t}.
\end{equation}
Such diffeomorphisms shift the hydrodynamic field by
\begin{equation}
\delta_H\xi_t=\delta T
=
c_0+c_+e^{r_0t}+c_-e^{-r_0t},
\end{equation}
which are precisely the horizon symmetry modes identified in the main text. The equivalence to the horizon symmetries discussion in Section \ref{sec:hor_sy} can be seen by evaluating $f(t) = \chi^{t}(t, r_0)$ and $Z(t) = \chi^{r}(t, r_0)$ for the diffeomorphisms~\eqref{eq:rad_JT_hor}. This gives precisely~\eqref{hongsymmetry} for the case $d=2$ with the identifications $a_1=c_0$, $a_2=c_-$ and $a_3 = - 2 r_0^2 c_+$. 

\subsection*{BTZ black hole}

\paragraph{} For BTZ we perform a similar analysis for a residual diffeomorphism $\delta g_{MN} \to \delta g_{MN} + \nabla_{M} \chi_{N} + \nabla_{N} \chi_{M}$ where
\begin{eqnarray}
\chi_r(t,r \to r_0, x) &=& \delta T(t,x) + \mathcal{O}((r-r_0)^2), \nonumber \\
\chi_x(t, r \to r_0, x)  &=&r^2 \delta X(t,x)+r\partial_x \delta T(t,x) + \mathcal{O}((r-r_0)^2), \nonumber   \\
\chi_t(t, r \to r_0, x) &=&-D(r)\delta T(t,x)-r\partial_t \delta T(t,x) + \delta R(t, x) + \mathcal{O}((r-r_0)^2),
\end{eqnarray}
such that they preserve radial gauge near the horizon. These diffeomorphisms act on the remaining metric components as
\begin{eqnarray}
\delta g_{tt}(r_0) &\to& \delta g_{tt}(r_0)   - 2 r_0 \delta R + 2 r_0^2 \partial_{t} \delta T + 2 \partial_{t} \delta R - 2 r_0 \partial_{t}^2 \delta T,\nonumber \\
\delta g_{tx}(r_0) &\to& \delta g_{tx}(r_0) + \partial_{x} \delta R + r_0^2 \partial_{t} \delta X, \nonumber \\
\partial_{r} \delta g_{tt}(r_0) &\to& \partial_{r} \delta g_{tt}(r_0)  - 2 \delta R - 2 \partial_{t}^2 \delta T.
\end{eqnarray}
The boundary conditions $\delta g_{tt}(r_0) = \delta g_{tx}(r_0) = \partial_{r} \delta g_{tt}(r_0) = 0$ are thus preserved by residual diffeomorphisms of the form 
\begin{equation}
\begin{split}
\delta T(t,x)
&=
b_0(x)+b_+(x)e^{r_0t}+b_-(x)e^{-r_0t},
\\
\delta R(t,x)
&=
-r_0^2b_+(x)e^{r_0t}
-r_0^2b_-(x)e^{-r_0t},
\\
\delta X(t,x)
&=
c_0(x)
+\frac{\partial_xb_+(x)}{r_0}e^{r_0t}
-\frac{\partial_xb_-(x)}{r_0}e^{-r_0t},
\end{split}
\end{equation}
which correspond to shifts of the hydrodynamic fields by
\begin{equation}
\begin{split}
\delta_H\xi_t = \delta T
&=
b_0(x)+b_+(x)e^{r_0t}+b_-(x)e^{-r_0t},
\\
\delta_H\xi_x = - \delta X
&=
-c_0(x)
-\frac{\partial_xb_+(x)}{r_0}e^{r_0t}
+\frac{\partial_xb_-(x)}{r_0}e^{-r_0t}.
\end{split}
\end{equation}
After redefining the arbitrary function $-c_0(x)$ as $c_x(x)$, these are precisely the zero modes of the BTZ action identified in the main text. To compare with the horizon symmetries in Section \ref{sec:hor_sy}, we use $f(t,x)=\chi^t(t,r_0,x)$, $Z(t,x)=\chi^r(t,r_0,x)$, $\tilde{X}(t,x)=\chi^x(t,r_0,x)$ to obtain
\begin{eqnarray}
f(t,x) &=& b_0(x) + b_+(x) e^{r_0 t} + b_{-}(x) e^{- r_0 t}, \nonumber \\
Z(t,x)&=&-2r_0^2 b_+(x)e^{r_0t}, \nonumber \\
 \widetilde X(t,x) &=&
c_0(x)
+\frac{1}{r_0}\partial_xb_0(x)
+\frac{2}{r_0}\partial_xb_+(x)e^{r_0t},
\end{eqnarray}
which are exactly of the form~\eqref{hongsymmetry} for $d=3$ after we identify $a_1(x)=b_0(x)$, $a_2(x)=b_-(x)$, $a_3(x)=-2r_0^2b_+(x)$ and $a_4^x(x)=c_0(x)+\frac{1}{r_0}\partial_x b_0(x)$.

\section{Dirichlet boundary conditions}\label{app:Dir}

\paragraph{}In this Appendix we derive an alternative form of the hydrodynamic action for JT gravity, by imposing different horizon boundary conditions. Specifically, we consider a more conventional Dirichlet prescription by introducing a timelike stretched horizon at $r=r_c$, adding a Gibbons--Hawking term on that surface,\footnote{Here we mean the usual Gibbons-Hawking term, not including the additional counterterms given in equation \eqref{eq:bd_JT}.} and holding fixed the induced data there. The steps of the calculation are analogous to those in Section \ref{sec:AdS2section} and so we will just report the differences when these boundary conditions are used. We also give a brief analysis of the results for Dirichlet boundary conditions for the BTZ case.

\subsection*{JT case}
\paragraph{}We use the same JT bulk action and asymptotic boundary term as in \eqref{ads2action} and \eqref{eq:bd_JT}, now on the regulated region $r_0\leq r_c\leq r\leq r_\Lambda$, and supplement them by a Gibbons--Hawking term at $r=r_c$. Its extrinsic curvature is evaluated using the outward-pointing normal of the regulated region, which points towards decreasing $r$ at the inner boundary. We denote the emblackening factor at the stretched horizon as
\begin{equation}
    D_c\equiv D(r_c)=r_c^2-r_0^2.
    \label{eq:JT_Dir_Dc}
\end{equation}

\subsection*{Partially on-shell solutions}

\paragraph{}We impose Dirichlet boundary conditions on the stretched horizon: fixing the induced metric perturbation $\delta g_{tt}(t,r_c)=0$ and the dilaton perturbation $\delta\Phi(t,r_c)=0$. For the partially on-shell solutions in the main text
\begin{equation}
\begin{split}
    \delta g_{tt}(t,r_c)
    ={}&-r_c\alpha_1(t)-\alpha_2(t)
      +2\partial_t\zeta^r(t,r_c)
      -2D_c\partial_t\zeta^t(t,r_c)
      -2r_c\zeta^r(t,r_c),
      \\
    \delta\Phi(t,r_c)
    ={}&a\zeta^r(t,r_c)+\frac{a}{2}\alpha_1(t)
       +\partial_t\alpha_4(t)+r_c\alpha_4(t).
    \label{eq:JT_Dir_fixed_data}
\end{split}
\end{equation}
Therefore, after imposing the same asymptotic boundary conditions as in the main text, $\alpha_1(t)$ and $\alpha_2(t)$ are now given by
\begin{equation}
\begin{split}
    \alpha_1(t)
    ={}&-2\zeta^r(t,r_c)
      -\frac{2}{a}(\partial_t+r_c)
      \left(
        \delta\phi^{(s)}(t)
        +\frac{a}{2}\delta g_{tt}^{(s)}(t)
        -a\partial_t\zeta_t^{(2)}(t)
      \right),
      \\
    \alpha_2(t)
    ={}&2\partial_t\zeta^r(t,r_c)
      -2D_c\partial_t\zeta^t(t,r_c)+\frac{2r_c}{a}(\partial_t+r_c)
      \left(
        \delta\phi^{(s)}(t)
        +\frac{a}{2}\delta g_{tt}^{(s)}(t)
        -a\partial_t\zeta_t^{(2)}(t)
      \right),
    \label{eq:JT_Dir_alpha12}
\end{split}
\end{equation}
and so the energy constraint \eqref{eq:bulk_energy} is
\begin{equation}
    {\rm energy}
    =\partial_t\left(
      D_c\left(
        2\delta\phi^{(s)}
        +a\left(\delta g_{tt}^{(s)}-2\partial_t\xi_t\right)
      \right)
    \right),
    \label{eq:JT_Dir_constraint}
\end{equation}
where, in direct analogy with \eqref{eq:hydro_variable_generic_JT}, the hydrodynamic variable
\begin{equation}
    \xi_t(t)
    =\big[-\zeta^t\big]_{r_c}^{\infty}
    =-\zeta^t(t,\infty)+\zeta^t(t,r_c),
    \label{eq:JT_Dir_cutoff_Wilson_line}
\end{equation}
is the relative time diffeomorphism between the stretched horizon and the asymptotic boundary. 

\subsubsection*{Hydrodynamic action} 

\paragraph{}Now we turn to the hydrodynamic action (written in terms of $\xi_t$ and the sources) found using these solutions. At linear order in amplitudes it is unchanged from \eqref{eq:linJT}, but at quadratic order it is different. The bulk contribution from the partially on-shell solutions (equation \eqref{eq:ene_mom}) can be written
\begin{equation}
    S_{\rm bulk,Dir}=\frac{D_c}{2}\int dt\,(\partial_t\xi_t)
    \left(      
        2\delta\phi^{(s)}
        +a\left(\delta g_{tt}^{(s)}-2\partial_t\xi_t\right)
    \right).
    \label{eq:JT_Dir_bulk_action}
\end{equation}
Direct substitution of the partially on-shell fields into the boundary term \eqref{boundary1}, together with the asymptotic boundary contribution \eqref{eq:bd_JT} and the new Gibbons--Hawking term at $r=r_c$, gives the following contribution to the quadratic action 
\begin{equation}
\begin{split}
    S_{\rm bdy,Dir}
    =\int dt\Bigg(&
      \frac{ar_0^2}{8}{\delta g_{tt}^{(s)}}^2
      -\frac{r_0^2}{2}\delta\phi^{(s)}\delta g_{tt}^{(s)}
      +\frac{1}{a}\delta\phi^{(s)}
       (\partial_t^2-r_0^2)\delta\phi^{(s)}
      \\
    &-\frac{D_c}{4a}
      \left(
        2\delta\phi^{(s)}+a\delta g_{tt}^{(s)}
      \right)
      \left(
        2\delta\phi^{(s)}
        +a\left(\delta g_{tt}^{(s)}-2\partial_t\xi_t\right)
      \right)
    \Bigg).
    \label{eq:JT_Dir_boundary_action}
\end{split}
\end{equation}
Combining \eqref{eq:JT_Dir_bulk_action} and \eqref{eq:JT_Dir_boundary_action} gives the quadratic hydrodynamic action with Dirichlet horizon boundary conditions as
\begin{equation}
\begin{aligned}
    S_{\rm hydro,Dir}^{(2)}
    =\int dt\Biggl(&\,
      \frac{ar_0^2}{8}{\delta g_{tt}^{(s)}}^2
      -\frac{r_0^2}{2}\delta\phi^{(s)}\delta g_{tt}^{(s)}
      +\frac{1}{a}\delta\phi^{(s)}
       (\partial_t^2-r_0^2)\delta\phi^{(s)}\\
       &\,-\frac{D_c}{4a}
      \left(
        2\delta\phi^{(s)}
        +a\left(\delta g_{tt}^{(s)}-2\partial_t\xi_t\right)
      \right)^2\Biggr).
    \label{eq:JT_Dir_action}
\end{aligned}
\end{equation}
For comparison with the action obtained in the main text, it is useful to integrate \eqref{hydroJT} by parts to obtain
\begin{equation}
\begin{split}
    S_{\rm hydro}^{(2)}
    =\int dt\Bigg(&
      \frac{ar_0^2}{8}{\delta g_{tt}^{(s)}}^2
      -\frac{r_0^2}{2}\delta\phi^{(s)}\delta g_{tt}^{(s)}
      +\frac{1}{a}\delta\phi^{(s)}
       (\partial_t^2-r_0^2)\delta\phi^{(s)}
      \\
    &-\frac{1}{4a}
      \left(
        2\delta\phi^{(s)}
        +a\left(\delta g_{tt}^{(s)}-2\partial_t\xi_t\right)
      \right)
      (\partial_t^2-r_0^2)
      \left(
        2\delta\phi^{(s)}
        +a\left(\delta g_{tt}^{(s)}-2\partial_t\xi_t\right)
      \right)
    \Bigg).
    \label{eq:JT_main_action_explicit_form}
\end{split}
\end{equation}
The difference is in the last term. We now have a second order, rather than fourth order, action for the hydrodynamic field $\xi_t$. On one hand this is advantageous: the fourth order Schwarzian action leads to exponentially growing modes which look unphysical, but ultimately are redundancies. On the other hand it has a drawback: the exponentially growing modes, invisible in \eqref{eq:JT_Dir_action}, directly reflect the four-point out-of-time-ordered correlators used to characterise quantum chaos. Notice that the action in~\eqref{eq:JT_Dir_action} is explicitly cut-off dependent through the factor of $D_{c} = r_c^2 - r_0^2$. From the point of view of a boundary hydrodynamic action this is somewhat unsatisfying, as there is no boundary interpretation of this cutoff parameter $r_c$. We note that similar cut-off dependence was seen in a previous attempt to construct hydrodynamic actions using Dirchlet boundary conditions near the horizon~\cite{deBoer:2015ija}. 

\paragraph{}In the absence of sources, the quadratic action with Dirichlet horizon conditions reduces to
\begin{equation}
    S^{(2)}_\text{hydro,Dir}[\xi_t]=-aD_c\int dt \left(\partial_t \xi_t\right)^2,
\end{equation}
which is manifestly different from the linearised Schwarzian action \eqref{hydroJTsourcefree} obtained in the main text. The limit in which the stretched horizon coincides with the event horizon $r_c=r_0$ corresponds to $D_c=0$. In this limit, $\xi_t$ drops out of the action completely and we will address it separately at the end of this Appendix.

\paragraph{}The hydrodynamic action obtained with these horizon boundary conditions is again diffeomorphism invariant. Combining the linear and quadratic terms it may be written as
\begin{equation}
    S_{\rm hydro,Dir}
    =\int dt\left(
      r_0^2\left(\varphi+\frac{a}{2}\mathcal B\right)
      +\frac{ar_0^2}{8}\mathcal B^2
      -\frac{r_0^2}{2}\varphi\mathcal B
      +\frac{1}{a}\varphi(\partial_t^2-r_0^2)\varphi-\frac{D_c}{a}
      \left(\varphi+\frac{a}{2}\mathcal B\right)^2
    \right),
    \label{eq:JT_Dir_action_pullback}
\end{equation}
to quadratic order in the perturbations, where $\mathcal{B}$ and $\varphi$ are the diffeomorphism-invariant fields defined in \eqref{eq:B_lin_quad} and \eqref{eq:curly_phi}. This can be compared to the action obtained with the horizon boundary conditions in the main text \eqref{B_cpt}, which after integration by parts is 
\begin{equation}
\begin{aligned}
S_{\mathrm{hydro}}
=
\int dt\Biggl(
r_0^2&\,\left(\varphi+\frac{a}{2}\mathcal B\right)+\frac{ar_0^2}{8}\mathcal B^2-\frac{r_0^2}{2}\varphi\mathcal{B}+\frac{1}{a}\varphi(\partial_t^2-r_0^2)\varphi\\
&\,-\frac{1}{a}\left(\varphi+\frac{a}{2}\mathcal B\right)\left(\partial_t^2-r_0^2\right)\left(\varphi+\frac{a}{2}\mathcal B\right)\Biggr).
    \label{eq:JT_Dir_vs_main_pullback}
\end{aligned}
\end{equation}
The difference is again in the final term, where $\partial_t^2$ in \eqref{eq:JT_Dir_vs_main_pullback} is replaced by $r_c^2$ in \eqref{eq:JT_Dir_action_pullback}.
In terms of the Weyl-invariant variable $\sigma(t)$ defined in \eqref{eq:sigmatdefn}, the action with Dirichlet horizon boundary conditions is (to quadratic order)
\begin{equation}
\label{eq:sigmatactionDirichlet}
        S_{\rm hydro,Dir}=\int dt\left(r_0^2(\sigma-a)-\frac{r_c^2}{a}(\sigma-a)^2+\frac{1}{a}\Psi\partial_t^2\Psi\right).
\end{equation}
To compare with equation \eqref{quadomegaaction} in the main text we need to remove the equilibrium term there, which corresponds to replacing $\sigma\rightarrow\sigma-a$. Upon doing this we see that the difference is just that $\partial_t^2$ in the quadratic in $(\sigma-a)$ part of the main text action is replaced by $r_c^2$. Since $\sigma(t)$ is Weyl-invariant, the action \eqref{eq:sigmatactionDirichlet} has the same Weyl transformation properties as the action obtained in the main text.

\subsubsection*{Two-point functions} 

\paragraph{}From the action, we now compute the two-point functions in the same way as in the main text. Defining the stress tensor and scalar expectation values as in \eqref{eq:vevs_T_phi}, we find explicitly that
\begin{equation}
\begin{split}
    \delta T_{\rm Dir}^{tt}
    ={}&ar_0^2\delta g_{tt}^{(s)}-r_0^2\delta\phi^{(s)}-D_c\left(
        2\delta\phi^{(s)}
        +a\left(\delta g_{tt}^{(s)}-2\partial_t\xi_t\right)
      \right),\\
    \delta\mathcal O_{\phi,{\rm Dir}}
    ={}&\frac{2}{a}(\partial_t^2-r_0^2)\delta\phi^{(s)}
      -\frac{D_c}{a}\left(
        2\delta\phi^{(s)}
        +a\left(\delta g_{tt}^{(s)}-2\partial_t\xi_t\right)
      \right).
    \label{eq:JT_Dir_constitutive_O}
\end{split}
\end{equation}
These satisfy the linearised Weyl Ward identity \eqref{eq:Ward_id_JT} identically (i.e.~off-shell). This is because the action \eqref{eq:sigmatactionDirichlet} has the same Weyl transformation properties as that in the main text \eqref{quadomegaaction}.

\paragraph{}To obtain the two-point functions we must integrate out the hydrodynamic field $\xi_t$. Its equation of motion is
\begin{equation}
\begin{aligned}
    0=\frac{\delta S_{\rm hydro,Dir}}{\delta\xi_t}&\,
    =-\partial_t\left(D_c\left(
        2\delta\phi^{(s)}
        +a\left(\delta g_{tt}^{(s)}-2\partial_t\xi_t\right)
      \right)
    \right)\\
    &\,=\partial_t\delta T_{\rm Dir}^{tt}
      -ar_0^2\partial_t\delta g_{tt}^{(s)}
      +r_0^2\partial_t\delta\phi^{(s)}
    =-\,{\rm energy}
    \label{eq:JT_Dir_action_variation}.
\end{aligned}
\end{equation}
The first equality on the second line shows that, just like in the main text, the hydrodynamic equation of motion is the linearisation of the conservation law $\nabla_t T^{tt}_\text{Dir}=\mathcal{O}_{\phi,\text{Dir}}\nabla^t\phi^{(s)}$. This is because the action is again diffeomorphism invariant. The last equality shows that integrating out the hydrodynamic field corresponds to imposing the remaining Einstein equation that we did not impose when deriving the action.

\paragraph{}For non-zero frequencies (and $D_c\ne0$) the solution to the hydrodynamic equation is
\begin{equation}
    \xi_t(\omega)
    =\frac{i}{2a\omega}
      \left(
        a\delta g_{tt}^{(s)}(\omega)
        +2\delta\phi^{(s)}(\omega)
      \right),
    \qquad \omega\neq0.
    \label{eq:JT_Dir_xi_onshell}
\end{equation}
Notice that for generic frequencies this agrees with the solution for the hydrodynamic field in the main text \eqref{eq:freq_2D}, obtained with different boundary conditions. The difference is that the solution for $\xi_t$ in the main text was not uniquely defined for $\omega=\pm ir_0$ due to the additional Schwarzian zero modes that can be added without without affecting the stress tensor two-point function. Putting this field on-shell yields and then using the same conventions as in the main text to obtain the response functions yields
\begin{equation}
\begin{aligned}
    \label{eq:JT_Dir_two_point_functions}
    G_{T^{tt}_{\text{Dir}}T^{tt}_{\text{Dir}}}(\omega)&\,=ar_0^2,
    \quad\quad\quad\quad\quad\quad
    G_{\mathcal O_{\phi,\text{Dir}}\mathcal O_{\phi,\text{Dir}}}(\omega)
    =-\frac{2}{a}(\omega^2+r_0^2),\\
    G_{T^{tt}_{\text{Dir}}\mathcal O_{\phi,\text{Dir}}}(\omega)
    &\,=G_{\mathcal O_{\phi,\text{Dir}} T^{tt}_{\text{Dir}}}(\omega)=-r_0^2,
\end{aligned}
\end{equation}
which are identical to those obtained using the mixed horizon boundary conditions (equation \eqref{eq:const_2D}). The reason is simple: these two-point functions are fixed by the Ward identities \eqref{eq:Ward_id_JT} and \eqref{eq:JT_Dir_action_variation}, which follow from the diffeormophism invariance and Weyl transformation properties of the action explained above.

\subsubsection*{Horizon symmetries}

\paragraph{}Since the horizon symmetries depend on the choice of boundary conditions, for the action we have just obtained they are different than those described in the main text.\footnote{Strictly speaking these are not horizon symmetries except when $r_c=r_0$. But since we have in mind $r_c$ being close to the horizon we will continue to use this phrase.} Consider a diffeomorphism $\chi(t,r)$ that vanishes near the asymptotic boundary but not on the stretched horizon, where
\begin{equation}
 \left.\chi\right|_{r=r_c}
   = f_c(t)\,\partial_t+Z_c(t)\,\partial_r .
\end{equation}
This diffeormorphism acts on the Dirichlet data on the stretched horizon as
\begin{equation}
 \left.\Delta_\chi\delta g_{tt}\right|_{r_c}
   =2\partial_t Z_c-2D_c\partial_t f_c-2r_cZ_c,
 \qquad
 \left.\Delta_\chi\delta\Phi\right|_{r_c}
   =aZ_c.
\end{equation}
We therefore have a horizon symmetry when $Z_c=0$ and $D_c\partial_t f_c=0$. Assuming $D_c\ne0$, the horizon symmetries with these boundary conditions are the time translations $f_c(t)=f_0$ for constant $f_0$. Under these the hydrodynamic field shifts
\begin{equation}
\label{eq:xitransformapp}
    \delta_H\xi_t=f_0,
\end{equation}
while the sources are invariant. The exponential horizon symmetries identified in the main text (equation \eqref{eq:ads2-xit-transform}) are no longer horizon symmetries with these boundary conditions.

\paragraph{}The hydrodynamic action \eqref{eq:JT_Dir_action} is manifestly invariant under the horizon symmetry transformation \eqref{eq:xitransformapp}, as are the constitutive relations \eqref{eq:JT_Dir_constitutive_O} and the hydrodynamic equation of motion \eqref{eq:JT_Dir_action_variation}, since $\xi_t$ appears in these only through $\partial_t\xi_t$. In other words, this is a zero mode that carries no energy. Note that in the absence of sources the hydrodynamic equation of motion \eqref{eq:JT_Dir_action_variation} also has a linear solution $\xi_t=f_1 t$ that does carry energy. This solution is not a horizon symmetry as, for $f_c=f_1t$, $\left.\Delta_\chi\delta g_{tt}\right|_{r_c}
   =-2D_c f_1 .$

\subsubsection*{The limit $r_c\rightarrow r_0$}

\paragraph{}So far we have implicitly assumed that the stretched horizon lies outside the event horizon $r_c>r_0$, such that $D_c>0$. The limit $r_c\rightarrow r_0$ corresponds to $D_c\rightarrow0$ for which the quadratic hydrodynamic action becomes
\begin{equation}
\begin{split}
    S_{\rm hydro,Dir}^{(2)}
    =\int dt\left(
      \frac{ar_0^2}{8}{\delta g_{tt}^{(s)}}^2
      -\frac{r_0^2}{2}\delta\phi^{(s)}\delta g_{tt}^{(s)}
      +\frac{1}{a}\delta\phi^{(s)}
       (\partial_t^2-r_0^2)\delta\phi^{(s)}\right).
    \label{eq:JT_Dir_horizon_action}
\end{split}
\end{equation}
In this limit $\xi_t$ drops out completely and so this does not choice does not generate a hydrodynamic action. Nevertheless, taking variations of \eqref{eq:JT_Dir_horizon_action} with respect to the sources does produce the same two-point functions as \eqref{eq:const_2D} and so \eqref{eq:JT_Dir_horizon_action} can be taken as the generating functional for stress tensor and scalar two-point functions directly, without first integrating out a hydrodynamic field.

\paragraph{}To understand why this happens, notice from equation \eqref{eq:JT_Dir_constraint} that when $D_c=0$ the `partially' on-shell solution in fact already satisfies the energy constraint. In other words, imposing Dirichlet conditions exactly at the horizon actually puts the gravitational action fully on-shell. It therefore produces the usual generating functional rather than a hydrodynamic action.

\subsection*{BTZ case}

\paragraph{}The same procedure can be carried out for the BTZ case, by
fixing all three components of the induced metric at $r=r_c$  (i.e. $\delta g_{\mu \nu}(r_c) = 0$) and
including the Gibbons--Hawking term there.\footnote{Again, at the horizon we do not include the counterterm in equation \eqref{btzaction}.} The linear action remains \eqref{eq:Slinearterm}. In the absence of sources,
the quadratic hydrodynamic action is
\begin{equation}
\begin{split}
S_{\mathrm{hydro,Dir}}^{(2)}
=\int d^2x\,\left(
\xi_t(\partial_t^2-\partial_x^2)(D_c+\partial_x^2)\xi_t
-2r_c\xi_t(\partial_t^2-\partial_x^2)\partial_x\xi_x
-r_c^2\xi_x(\partial_t^2-\partial_x^2)\xi_x
\right),
\end{split}
\label{eq:BTZ_Dir_sourcefree}
\end{equation}
which is quadratic in temporal derivatives. This is different from the action \eqref{eq:BTZ-action-nosources} obtained in the main text, which was quartic in temporal derivatives. 

\paragraph{} Writing $q_c=\partial_x\xi_t-r_c\xi_x$, the additional source terms at quadratic order are (up to total derivatives)
\begin{align}
 \Delta S^{(2)}_{\mathrm{hydro,Dir}}
 =\int d^2x\,\Bigl(&
 \left(\delta g^{(s)}_{tt}+\delta g^{(s)}_{xx}\right)
 \left(D_c\partial_t\xi_t
       +(\partial_t-r_c)\partial_xq_c\right)
 \nonumber\\
 &+2\delta g^{(s)}_{tx}
 \left((r_c\partial_t-\partial_x^2)q_c
       -D_c\partial_x\xi_t\right)
 +\mathcal L_{\mathrm{source,Dir}}\Bigr),
 \label{eq:BTZ_Dir_UV_sources}
\end{align}
where
\begin{align}
 \mathcal L_{\mathrm{source,Dir}}={}&
 \frac{3r_0^2-2r_c^2}{8}
       {\delta g^{(s)}_{tt}}^2
 -\frac{r_0^2+2r_c^2}{8}
       {\delta g^{(s)}_{xx}}^2
 +\frac{r_0^2-2r_c^2}{4}
       \delta g^{(s)}_{tt}\delta g^{(s)}_{xx}
 \nonumber\\
 &+\frac{2r_c^2-r_0^2}{2}
       {\delta g^{(s)}_{tx}}^2
 -\frac14\delta g^{(s)}_{tt}
       \partial_x^2\delta g^{(s)}_{tt}
 +\frac14\delta g^{(s)}_{xx}
       \partial_t^2\delta g^{(s)}_{xx}
 \nonumber\\
 &+\delta g^{(s)}_{tx}
       \partial_x^2\delta g^{(s)}_{tx}
 +r_c\delta g^{(s)}_{tt}
       \partial_x\delta g^{(s)}_{tx}
 +\delta g^{(s)}_{xx}
       (r_c-\partial_t)\partial_x\delta g^{(s)}_{tx}.
 \label{eq:BTZ_Dir_UV_contact}
\end{align}
This action is again cutoff dependent. At $r_c=r_0$, the complete quadratic action (including source terms) depends on the hydrodynamic fields only
through $q_c$, as we will see below.

\paragraph{}Using the definition \eqref{eq:BTZ_st_defn}, the linearised
constitutive relations for the stress tensor in terms of the hydrodynamics fields are
\begin{equation}
\begin{aligned}
 \delta T^{tt}_{\mathrm{Dir}}
 ={}&2D_c\partial_t\xi_t
     +2(\partial_t-r_c)\partial_xq_c
     +(2r_0^2-r_c^2-\partial_x^2)\delta g^{(s)}_{tt}-r_c^2\delta g^{(s)}_{xx}
     +2r_c\partial_x\delta g^{(s)}_{tx},\\[2pt]
 \delta T^{xx}_{\mathrm{Dir}}
 ={}&2D_c\partial_t\xi_t
     +2(\partial_t-r_c)\partial_xq_c
     -D_c\delta g^{(s)}_{tt}+(\partial_t^2-r_c^2-r_0^2)\delta g^{(s)}_{xx}
     +2(r_c-\partial_t)\partial_x\delta g^{(s)}_{tx},\\[2pt]
 \delta T^{tx}_{\mathrm{Dir}}
 ={}&2(r_c\partial_t-\partial_x^2)q_c
     -2D_c\partial_x\xi_t
     +(2r_c^2-r_0^2+2\partial_x^2)\delta g^{(s)}_{tx}-r_c\partial_x\delta g^{(s)}_{tt}
     -(r_c+\partial_t)\partial_x\delta g^{(s)}_{xx},
\end{aligned}
\label{eq:BTZ_Dir_constitutive}
\end{equation}
with Dirichlet boundary conditions. These are again cutoff dependent. However after integrating out the hydrodynamic fields $\xi_t, \xi_x$ using their equations of motion, one obtains (for generic $\omega, k$) the same two point functions presented in~\eqref{eq:ads32ptfns}.

\paragraph{}From the point of view of computing stress tensor two-point functions this action is therefore equally as good as the one presented in the main text. However it has different horizon symmetries. In this case they are the diffeomorphisms that preserve the Dirichlet
conditions $\delta g_{\mu\nu}(r_c)=0$, with
$\mu,\nu\in\{t,x\}$ (as before we have in mind $r_c$ close to $r_0$ and so still call these horizon symmetries). To determine these explicitly, consider a diffeomorphism that vanishes near the asymptotic boundary and at the stretched horizon is
\begin{equation}
 \left.\chi\right|_{r=r_c}
 =f_c(t,x)\partial_t+Z_c(t,x)\partial_r
  +\widetilde X_c(t,x)\partial_x .
\end{equation}
Its action on the induced metric is
\begin{equation}
\begin{split}
 \left.\Delta_\chi\delta g_{tt}\right|_{r_c}
 &=2\partial_tZ_c-2D_c\partial_tf_c-2r_cZ_c,\\
 \left.\Delta_\chi\delta g_{tx}\right|_{r_c}
 &=\partial_xZ_c-D_c\partial_xf_c
   +r_c^2\partial_t\widetilde X_c,\\
 \left.\Delta_\chi\delta g_{xx}\right|_{r_c}
 &=2r_cZ_c+2r_c^2\partial_x\widetilde X_c.
\end{split}
\end{equation}
The diffeomorphisms that don't change the induced metric are
\begin{align}
 \widetilde X_c&=F(x+t)+G(x-t), \nonumber \\
 Z_c&=-r_c\bigl(F'(x+t)+G'(x-t)\bigr),\nonumber \\
 f_c&=f_0+\frac{r_c^2}{D_c}\bigl(F(x+t)-G(x-t)\bigr])
          -\frac{r_c}{D_c}\bigl(F'(x+t)+G'(x-t)\bigr),
          \label{horizonsymdirichletads3}
\end{align}
where $F$ and $G$ are arbitrary functions, primes denote derivatives with respect to the arguments, and $f_0$ is a constant. Near-horizon diffeomorphisms transform the hydrodynamic fields as
\begin{equation}
 \delta_H\xi_t=f_c,
 \qquad\qquad
 \delta_H\xi_x=-\widetilde X_c
              +\frac{1}{r_c}\partial_xf_c.
\end{equation}
One  can verify that for the specific diffeomorphisms in~\eqref{horizonsymdirichletads3} these transformations are symmetries of the equation of motions, constitutive relations and action.

\paragraph{}As for the case of JT gravity, the action is singular as we take the cutoff surface to the horizon $r_c\to r_0$. In this case the source-free action reduces, after
integration by parts, to
\begin{equation}
S_{\mathrm{hydro,Dir}}^{(2)}
=-\int d^2x\,
(\partial_x\xi_t-r_0\xi_x)
(\partial_t^2-\partial_x^2)
(\partial_x\xi_t-r_0\xi_x),
\label{eq:BTZ_Dir_horizon}
\end{equation}
which can be viewed as a functional of a single field $q_c = \partial_{x} \xi_t-r_0 \xi_x$. As a result, the equations of motion obtained by varying the action with respect to $\xi_t, \xi_x$ are no longer independent
 \begin{equation}
 \mathrm{momentum} = -\frac{\delta S_{\mathrm{hydro,Dir}}^{(2)}}{\delta \xi_x},\quad\quad\quad\quad  \mathrm{energy} = -\frac{\delta S_{\mathrm{hydro,Dir}}^{(2)}}{\delta \xi_t} = -\frac{1}{r_0} \partial_{x}    \frac{\delta S_{\mathrm{hydro,Dir}}^{(2)}}{\delta \xi_x} .
 \end{equation} 
In other words, the linear combination $ r_0 \mathrm{energy} - \partial_{x} \mathrm{momentum}$ of the Einstein constraint equations is automatically satisfied by any field configuration. There is therefore only one non-trivial constraint equation, which is the equation of motion for the field  $q_c = \partial_{x} \xi_t-r_0 \xi_x$. Nevertheless, after including external sources and integrating out this field using its equation of motion, one still obtains the correct two-point functions \eqref{eq:ads32ptfns} for the stress tensor. 

\paragraph{} For this special case $r_c = r_0$ the set of horizon symmetries becomes 
\begin{equation}
\begin{aligned}
\label{eq:BTZ_hor}
 \widetilde X_c&= A e^{r_0(t + x)} + B e^{r_0(t - x)} + C,\quad\quad\quad\quad Z_c=-r^2_0 A e^{r_0(t + x)} + r_0^2 B e^{r_0(t - x)}, \\
 f_c&= f(t,x),
\end{aligned}
\end{equation}
with $A, B, C$ arbitrary constants. These generate shifts of the field $q_c =  \partial_{x} \xi_t-r_0 \xi_x$ as $\delta_{H} q_c = r_0 \tilde{X}_c$. The exponential form of the horizon symmetry that act on $q_c$ precisely matches the exponential form appearing in the OTOC. We emphasise this is a different set of horizon symmetries to those proposed in~\cite{Knysh:2024asf} and obtained in the main text from mixed horizon boundary conditions. The horizon symmetries of~\cite{Knysh:2024asf} include exponential modes with arbitrary spatial profile, while for the Dirichlet boundary conditions here the spatial profile is fixed.

\paragraph{} Finally we compute the two-point functions of the hydrodynamic
fields, as in \eqref{eq:BTZ_xit_xit}. Integrating \eqref{eq:BTZ_Dir_sourcefree} by parts gives
\begin{equation}
 S_{\mathrm{hydro,Dir}}^{(2)}
 =\int d^2x\,
 \left(
 D_c\,\xi_t(\partial_t^2-\partial_x^2)\xi_t
 -q_c(\partial_t^2-\partial_x^2)q_c
 \right),
 \label{eq:BTZ_Dir_diagonal_action}
\end{equation}
and so the quadratic kernel in the
$(\xi_t,\xi_x)$ basis is
\begin{equation}
 K_{\mathrm{Dir}}(\omega,k)
 =(\omega^2-k^2)
 \begin{pmatrix}
 k^2-D_c & ir_ck\\
 -ir_ck & r_c^2
 \end{pmatrix},
 \qquad
 \det K_{\mathrm{Dir}}
 =-D_cr_c^2(\omega^2-k^2)^2 .
 \label{eq:BTZ_Dir_kernel}
\end{equation}
For $r_c>r_0$, its inverse gives
\begin{equation}
\begin{aligned}
 \langle\xi_t(\omega,k)\xi_t(-\omega,-k)\rangle_{\mathrm{Dir}}
 &=-\frac{1}{D_c(\omega^2-k^2)},\\ 
 \langle\xi_x(\omega,k)\xi_x(-\omega,-k)\rangle_{\mathrm{Dir}}
 &=\frac{D_c-k^2}{r_c^2D_c(\omega^2-k^2)},\\ 
 \langle\xi_t(\omega,k)\xi_x(-\omega,-k)\rangle_{\mathrm{Dir}}
 &=-\langle\xi_x(\omega,k)\xi_t(-\omega,-k)\rangle_{\mathrm{Dir}}=\frac{ik}{r_cD_c(\omega^2-k^2)}.
\end{aligned}
\label{eq:BTZ_Dir_hydro_correlators}
\end{equation}

\bibliography{HydroActionPaper}
\bibliographystyle{JHEP}

\end{document}